\documentclass[twocolumn,aps,superscriptaddress,showpacs]{revtex4}
\usepackage{amsmath}
\usepackage{amssymb}
\usepackage{graphicx}

\begin{document}
\title{Few-Body Bound Complexes in One-dimensional Dipolar Gases and Non-Destructive Optical Detection}
\author{N.~T. Zinner}
\affiliation{Department of Physics and Astronomy, Aarhus University,  Aarhus C, DK-8000}
\affiliation{Department of Physics, Harvard University, Cambridge MA, 02138}
\author{B. Wunsch}
\affiliation{Department of Physics, Harvard University, Cambridge MA, 02138}
\affiliation{ABB Switzerland Ltd., Corporate Research}
\author{I. B. Mekhov}
\affiliation{Department of Physics, Harvard University, Cambridge MA, 02138}
\affiliation{University of Oxford, Department of Physics, Clarendon Laboratory, Parks Road, Oxford OX1 3PU, UK}
\author{S.-J. Huang}
\author{D.-W. Wang}
\affiliation{Physics department and NCTS, National Tsing-Hua University, Hsinchu 300, Taiwan}
\affiliation{Frontier Research Center on Fundamental and Applied Sciences of Matters, National Tsing-Hua University, Hsinchu 300, Taiwan }
\author{E. Demler}
\affiliation{Department of Physics, Harvard University, Cambridge MA, 02138}

\date{\today}

\date{\today}
\begin{abstract}
We consider dipolar interactions between heteronuclear molecules in low-dimensional geometries.  The setup consists of two one-dimensional tubes. We study the stability of possible few-body complexes in the regime of repulsive intratube interaction, where the binding arises from intertube attraction. The stable dimers, trimers, and tetramers are found and we discuss their properties for both bosonic and fermionic molecules. To observe these complexes we propose an optical non-destructive detection scheme that enables in-situ observation of the creation and dissociation of the few-body complexes. A detailed description of the expected signal of such measurements is given using the numerically calculated wave functions of the bound states.  We also discuss implications on the many-body physics of dipolar systems in tubular geometries, as well as experimental issues related to the external harmonic confinement along the tube and the prospect of applying an in-tube optical lattice to increase the effective dipole strength.  
\end{abstract}
\maketitle

\section{Introduction}
Ultracold polar molecules with anisotropic long-range interactions have generated a lot of interest recently. In recent experiments heteronuclear molecules could be  prepared in their rotational and vibrational  ground-state and cooled to  temperatures close
to quantum degeneracy \cite{doyle2004,ospelkaus2008,ni2008,deiglmayr2008,lang2008,ospelkaus2010,ni2010}, where
many exotic many-body states have been predicted \cite{baranov2008,lahaye2009}. In a three-dimensional
sample one faces the problem that attractive head-to-tail interactions of the dipoles can destabilize the system by strong particle loss~\cite{lushnikov2002}.
In order to overcome this problem, molecules can either be dressed
by ac-external fields leading to strongly repulsive interactions at small inter-particle distances or optical lattices such that particles cannot approach each other in the head-to-tail direction \cite{wang2006,wang2007,micheli2007,gorshkov2008}. 
Recently, experiments have entered the quasi-two-dimensional regime by applying an optical lattice potential to the
three-dimensional sample and observed large geometric effects on the loss rates \cite{miranda2011}. While this system 
does not provide the idealized stack of two-dimensional traps, it spurs hope that this can be realized in the near
future.
In the meantime a number of theoretical works on layered dipolar systems have appeared. For a single layer of dipolar molecules, $p$-wave superfluids \cite{bruun2008,cooper2009} and density-waves \cite{yamaguchi2010,sun2010,zinner2011,babadi2011} have been predicted. For large dipole moments and perpendicular dipoles a single layer is expected to
crystallize in a triangular array \cite{baranov2005,buchler2007,baranov2008,lu2008,baranov2008b,deu2010,cremon2010,tian2010} analogous to the famous Wigner crystal in the Coulomb gas \cite{wigner1934,bonsall1977}.
For bilayers or more generally multilayers, the long-range character of dipolar forces causes interlayer interactions that can lead to interlayer coherence and ferromagnet-type states \cite{dsarma2009}, BCS-BEC crossover phenomena \cite{pikovski2010,zinner2010} for bilayers, and dimerized pairing \cite{potter2010} for multilayers. Interlayer interactions have been studied extensively for analogue systems with long-range Coulomb interaction in quantum Hall \cite{spielman2000} or graphene bilayers \cite{novoselov2006} and
in electron-hole semiconductor bilayers where exciton superfluids or supersolids might be realized \cite{eisenstein2004,ye2010}. The few-body physics in single- or bilayer systems have also been studied recently \cite{shih2009,jeremy2010,klawunn2010,volosniev2010,volosniev2011,armstrong2011,volosniev2011b}, and 
some aspects of the many-body problem in coupled one-dimensional geometries have been discussed \cite{kollath2008,huang2009,chang2009,burovski2009,sudan2009,Dalmonte}. The details of Efimov physics for three dipolar bosons in three dimensions have also been 
discussed \cite{wang2011}.
In a recent letter \cite{letter} we predict the stability of complex bound states of dipolar molecules, that consist of more than one molecules per tube. In the present work we provide the details of our results and elaborately analyse the full parameter space wherein complexes are stable. 

\begin{figure}[ht!]
\begin{center}
\includegraphics[width=0.9\linewidth,angle=0]{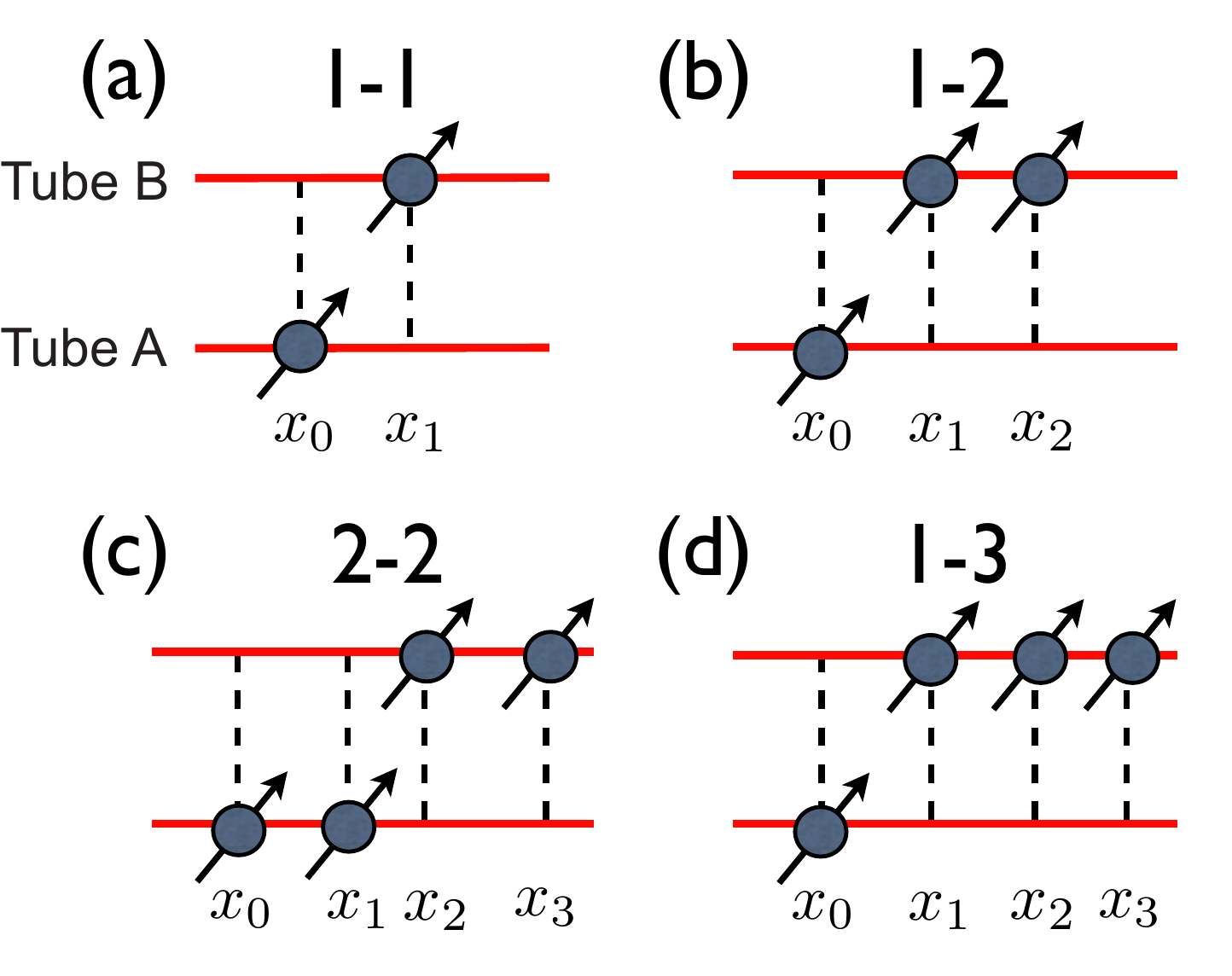}
\caption{(Color online)(a)-(d) Schematic pictures of the few-body complexes labeled by the number of molecules in the tubes $A$ and $B$. (a) 1-1 dimer (b) 1-2 trimer,
(c) 2-2 four-body state and (d) 1-3 four-body state. Also indicated are the coordinates of
the individual particles.}
\label{Fig:Complexes}
\end{center}
\end{figure}

Our basic setup consist of two parallel one-dimensional tubes at a fixed distance containing
ultracold polar molecules. While the motion is only along the tubes, the long-range character
of the dipole-dipole interaction means that the molecules interact across the tubes.
We consider which few-body bound states for the bi-tube system are stable as
the direction and the strength of the dipoles are varied. 
We focus on the regime where intratube interactions are repulsive so that the binding interactions of the few-body states comes from the intertube attractions. The formation of a trimer in our setup strongly resembles the formation of charged excitons in semiconducting nanotubes \cite{Trion} which are also 1D structures. A positively charged exciton consists of two holes that are bound together by an additional conduction electron. Analogously, in our setup two molecules in the same tube repel each other, but can be bound together by the attraction from a molecule in the other tube.

\begin{figure}[t!]
\begin{center}
\includegraphics[width=0.9\linewidth,angle=0]{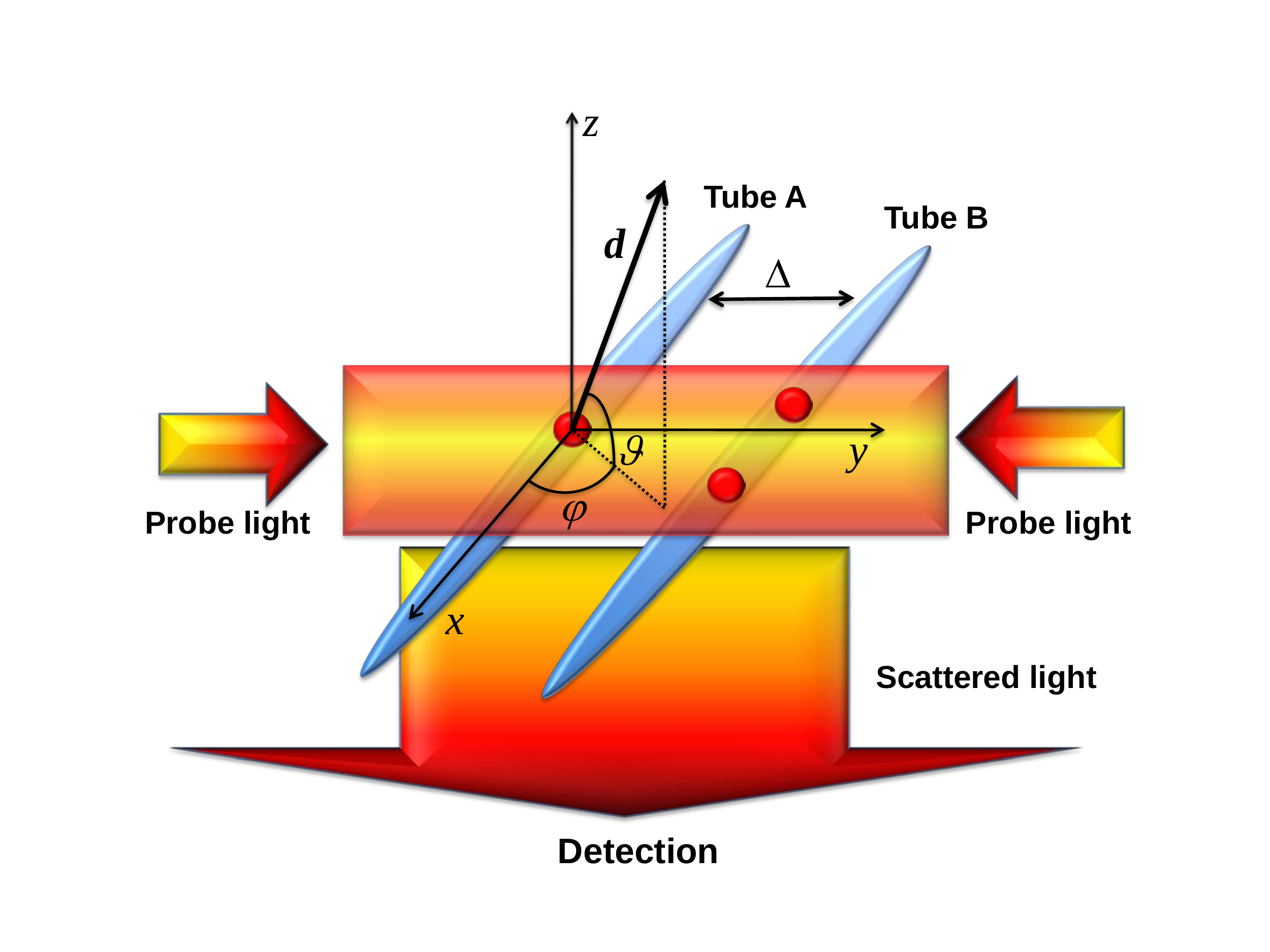}
\caption{(Color online) Visualization of setup. Molecules move in two parallel tubes named $A$ and $B$ that are located in $xy$-plane and are separated by distance $\Delta$. Red circles illustrate dipolar molecules. The  dipole moment is represented by an arrow ${\bf d}=d(\cos(\vartheta) \cos(\varphi),\cos(\vartheta) \sin(\varphi),\sin(\vartheta))$. $\varphi$  denotes the angle between tube and dipole within $xy$-plane and $\vartheta$ the angle out of the plane. The optical detection scheme consists of an additional probe beam and a detector and is discussed in Sec.\ref{Sec:Detection}. }
\label{Fig:Setup}
\end{center}
\end{figure}

We label the few-body complexes by the numbers $N_A$-$N_B$ of particles in tubes $A$ and $B$ that participate in the complex as shown in Fig.\ref{Fig:Complexes}. Below we discuss the stability of the 1-1, 1-2, 2-2 and 1-3 complexes in detail.
Then we propose an optical measurement scheme to detect various complexes for which
a schematic picture of the corresponding experimental setup is shown in Fig.~\ref{Fig:Setup}.
We show that the light scattering is strikingly different for all considered few-body states. Since it is non-destructive it allows the in-situ observation of the creation and dissociation of the few-body complexes.
We also discuss various extensions of our work including the case of parallel layers instead of tubes, multi-layer or multi-tube systems and effects of few-body bound states in the many body problem corresponding to a finite particle density. Technical details of the calculations are given in the Appendix.

\section{Model}\label{Model}
The general Hamiltonian for $N$ dipoles is given by
\begin{align}
H=\frac{1}{2m}\sum_{j=0}^{N-1} {\bf p}_j^{\,2}+\sum_{i<j} V_\text{d}({\bf r}_i-{\bf r}_j),
\end{align}
where $i,j$ label the particle and run from $0$ to $N-1$. We assume that all dipoles are aligned by an external electric field and describe strength and direction of the dipoles by the vector
\begin{align}
{\bf d}&=d(\cos(\vartheta) \cos(\varphi),\cos(\vartheta) \sin(\varphi),\sin(\vartheta)).
\end{align}
A schematic picture of the corresponding experimental setup is shown in Fig.~\ref{Fig:Setup}.
The interaction between two dipolar molecules is given by
\begin{align}
V_\textrm{d}({\bf r})=\frac{D^2}{r^3}(1-3\cos^2 \varphi_{rd}),\label{eq:dipole}
\end{align}
with $D^2=d^2/4\pi \epsilon_0$ and $\cos \varphi_{rd}={\bf r}\cdot {\bf d}/(r d)$, where $\bf r$ denotes the relative position ${\bf r}_{1}-{\bf r}_{2}$ of the two molecules. The motion of the molecules is reduced to two 1D tubes at positions $(x,y,z)=(x,\pm \Delta/2,0)$ where
$\Delta$ is the intertube distance. We consider deep 1D optical lattices where intertube tunneling can be negelected and where the transverse confinement length, $l_\perp$, is much smaller than the distance $\Delta$ between the tubes. In the limit $\Delta\gg l_\perp$
the intertube interaction, $V_1(x)$, and the intratube interaction, $V_0(x)$,
depend only on the distance between the particles along the tubes in the $x$-direction. However, we take into account that the intratube interaction
is modified at small distances by the transverse part of the wave function \cite{deu2010}
\begin{align}
V_0(x)&=\frac{D^2}{\Delta^3}(1-3 \cos^2 \varphi \cos^2 \vartheta)\lambda^3 f_0(\lambda \tilde{x})\label{eq:Vintra},\,\,\text{where}\\
f_0(u)&=\frac{-2 u +\sqrt{2 \pi} \left(1+u^2\right)\exp(u^2/2)\text{Erfc}\left(u/\sqrt{2}\right )}{4},\notag\\
\lambda&=\Delta/l_\perp \label{lambda}
\end{align}
with $\tilde{x}=x/\Delta$, and where $l_\perp$ is the transverse confinement length. In the strict 1D limit one replaces $\lambda^3 f_0(\lambda \tilde{x})\to 1/\tilde{x}^3$ in Eq.~\eqref{eq:Vintra}.
The finite transverse confinement reduces the intratube repulsion at small distances (of order $l_\perp$) and regularizes the $1/x^3$ behavior in the strict 1D case \cite{deu2010}.
Since $V_0(0)=(1-3 \cos^2 \varphi \cos^2 \vartheta)\frac{D^2}{l_\perp^3} \sqrt{\frac{\pi}{8}}$ and the radius of the core is of order $l_\perp$, the short range behaviour of the intratube interaction is similar to a contact interaction $ \delta(x) D^2/l_\perp^2$.
The short range part is particularly important for bosonic molecules.
For $\lambda\gg1$ this short-range repulsion exceeds intertube attraction and bosons become hard-core and behave as fermions in 1D. However, this is different at tilting angles where the prefactor $(1-3 \cos^2 \varphi \cos^2 \vartheta)=0$. For $\vartheta=0$, this condition defines the so-called magic angle, $\varphi_M=\arccos(\frac{1}{\sqrt{3}})$.

\begin{figure}[t!]
\begin{center}
\includegraphics[width=0.8\linewidth,angle=0]{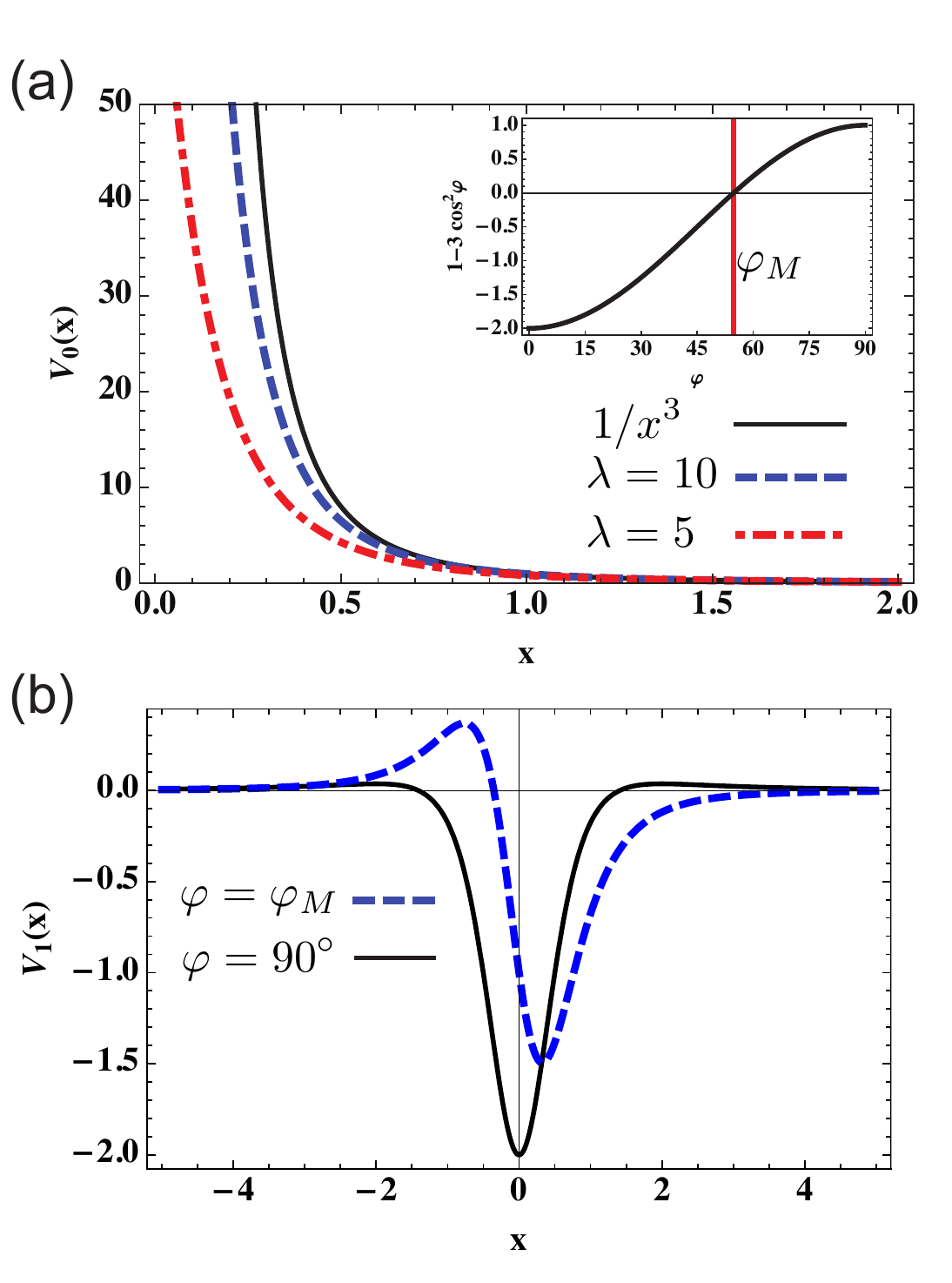}
\caption{(Color online) (a) Main panel shows the intratube potential $V_0(x)$ for perpendicular dipoles $\varphi=90^\circ, \vartheta=0^\circ$ for various strengths of the transverse confinement $\lambda$ see Eq.~\eqref{eq:Vintra}. Tilting the dipoles leads to a suppression of $V_0$ without changing its functional form.  As shown in the inset the intratube interaction vanishes at the magic angle $\varphi_M$.  (b) The intertube interaction for $\varphi=\varphi_M$ and $\varphi=90^\circ$ both with $\vartheta=0^\circ$. Since particles in different tubes are always separated by at least the intertube distance $\Delta$ the dependence of $V_1(x)$ on the transverse confinement is of order $1/\lambda^2$ and is neglected in this paper. Energy in units of $\hbar^2/m\Delta^2$ and length in units of
$\Delta$.}
\label{Fig:Potentials}
\end{center}
\end{figure}

Corrections
to the intertube interaction in quasi-1D scale with $1/\lambda^2$ and as we have used $\lambda=5$ and $10$ in our calculations these
can be neglected. In turn, the intertube interaction is
\begin{align}
V_1(x)&=\frac{D^2}{\Delta^3}\frac{\left(1-3\frac{\cos(\vartheta)^2\left(\tilde{x} \cos(\varphi)+\sin(\varphi)\right)^2}{\tilde{x}^2+1}\right)}{(\tilde{x}^2+1)^{3/2}}
\end{align}
with $\tilde{x}=x/\Delta$.
Figure~\ref{Fig:Potentials} shows the intratube and intertube interaction and illustrates their dependence on tilting angle and transverse confinement.

The characteristic energies for dipole interaction and kinetic energy are $D^2/\Delta^3$ and $\hbar^2/m \Delta^2$ respectively which defines a dimensionless measure for the interaction strength given by
\begin{align}
U_0=m D^2/\Delta \hbar^2.
\end{align}
$U_0$ determines the competition of interaction to kinetic energy.  The classical limit where the kinetic energy of the bound state can be neglected corresponds to $U_0\to \infty$.
If not stated, otherwise we measure length in units of $\Delta$ and energies in
units of $\hbar^2/m\Delta^2$.

We consider few-particle complexes of up to four molecules in total as shown pictorially in Fig.~\ref{Fig:Complexes}. We characterize the configurations by the number of molecules in tubes $A$ and $B$. For example the four body complex 1-3 consists of 1 molecule in tube $A$ and 3 in  tube $B$.
The interaction potentials for the various complexes are given by:
\begin{eqnarray}
V_{1-1}&=&V_1(x_{10})\\
V_{1-2}&=&V_1(x_{10})+V_1(x_{20})+V_0(x_{21})\label{eq:trimer}\\
V_{2-2}&=&V_1(x_{20})+V_1(x_{30})+V_1(x_{21})+V_1(x_{31})\notag\\
& &+V_0(x_{10})+V_0(x_{32})\\
V_{1-3}&=&V_1(x_{10})+V_1(x_{20})+V_1(x_{30})\notag\\
& &+V_0(x_{21})+V_0(x_{31})+V_0(x_{32})
\end{eqnarray}
with $x_{ij}=x_i-x_j$.
Since the interaction only depends on the relative coordinates, their center of mass motion can be separated.
To a given set of parameters $\vartheta,\varphi,\lambda,U_0, N_A, N_B$ we determine the eigenspectrum by solving the Hamiltonian of the relative motion numerically using a finite difference method which turns the solution of the Schr\"odinger equation into a diagonalization problem of large sparse matrices as discussed in Appendix~\ref{RelCoord}. The stability of a few -body complex is checked in two complementary ways. First we confirm that the eigenenergies of the $N= N_A+N_B$ particle state is smaller than any other smaller few-body state, so that there is an energy penalty against dissociation. For example a stable 1-2 state has to have smaller energy than the 1-1 state. This argument is strictly true only in the zero density limit and in absence of an external parabolic confinement potential, so that it is always possible to have a free particle with zero energy.
The second method is to check the average distance between the particles. In a bound state all distances have to be finite.

\section{Few-body bound states }\label{stability}
In this section we work in the low density limit and discuss how few-body states consisting of a small number of particles in each tube depend on the system parameters $U_0$, $\varphi$, $\vartheta$ and $\lambda$ as well as on whether the molecules are fermions or bosons. Effects related to a finite density per tube will be addressed in Sec. \ref{many-body}. We focus on the regime where the intratube interaction
is repulsive so that few-body complexes are bound due to the intertube attraction. In the case of intratube attraction higher transverse bands or even  chemical reactions might become important and the effective intratube interaction given in Eq.~\eqref{eq:Vintra} should be modified accordingly at small length scales.

\begin{figure}[t!]
\begin{center}
\includegraphics[width=0.9 \linewidth,angle=0]{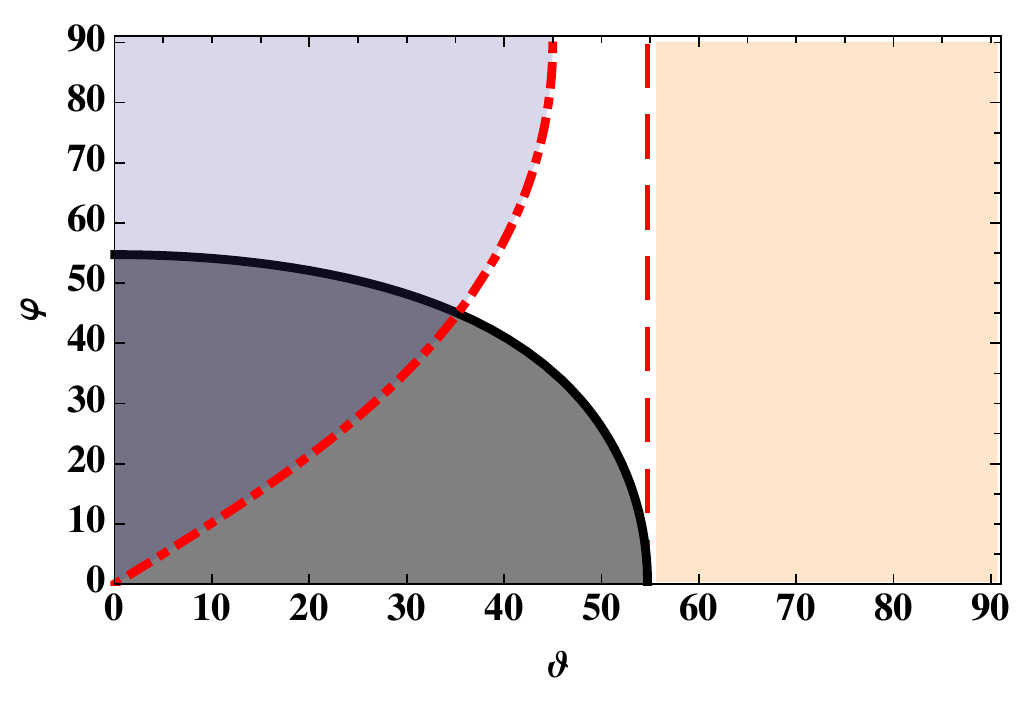}
\caption{(Color online) Different regimes for few-body states in tubes. In the left upper corner above the dashed-dotted line $\int\,dx V_1(x)<0$ and there is always a bound dimer state. In the center of the figure between dashed-dotted and dashed line there is intertube attraction but $\int\,dx V_1(x)>0$, so that a dimer only exists above a critical interaction strength. To the right of the (red) dashed line $V_1(x)>0$ everywhere. In the left lower corner limited by the solid (black) line, the intratube interaction is repulsive.}
\label{DimerRegimes}
\end{center}
\end{figure}

The angles $\vartheta$ and $\varphi$ introduced in Fig.\ref{Fig:Setup} will determine the ratio between intratube and intertube interaction. For a given $\varphi$ the intertube attraction $V_1$ is strongest for $\vartheta=0$ and in order to stabilize larger complexes we will mainly focus on  $\vartheta=0$ in this section. An exemption is the dimer which turns out to be always stable at $\vartheta=0$, but can be made unstable at $\vartheta>0$ as shown in Fig.~\ref{DimerRegimes}. The intratube interaction $V_0$ becomes repulsive for  $\varphi>\arccos(\frac{1}{\sqrt{3}\cos(\vartheta)})$.  At $\vartheta=0$ the intratube interaction becomes repulsive for $\varphi>\varphi_M$.

\subsection{Dimer}\label{sec:dimer}

We start our discussion with the simplest few-body state, the dimer, that consists of one particle per tube as sketched in Fig.~\ref{Fig:Complexes}(a).
The stability of the dimer as function of the angles $\vartheta$ and $\varphi$ is shown in Fig.~\ref{DimerRegimes}. For $\vartheta=0$ a dimer can always profit from the attractive part of the intertube interaction and is stable for all $\varphi$ and $U_0$. This can be shown by noting that $\int V_{1-1}(x) dx<0$, which is sufficient to ensure a bound state in 1D. Even in two dimensional systems where $\int V_{1-1}(x) dx=0$ a dimer always exists \cite{volosniev2010}. However, in 1D the dimer can be made unstable by rotating the dipoles out of the $x-y$ plane corresponding to $\vartheta>0$.  For general $\vartheta$ the integral over the intertube interaction is $\int\,dx V_1(x)=2 \left(\cos^2(\varphi)\cos^2(\vartheta)-\cos(2 \vartheta)\right)D^2/\Delta^3$.
There are three regimes: 1)  For $\int\,dx V_1(x)<0$ there is a dimer bound state for any $U_0$. This regime is in the upper left corner and is limited from below by dashed dotted line in Fig.~\ref{DimerRegimes}, which corresponds to $\varphi=\arccos(\sqrt{\frac{\cos(2\vartheta)}{\cos^2(\vartheta)}})$. 2) For $\int\,dx V_1(x)>0$ but $V_1(x_{min})<0$ there is a bound dimer above a critical interaction strength. Here $x_{min}$ is the position of the minimum of $V_1{x}$ around $x=0$ (as seen for instance in Fig.~\ref{Fig:Potentials}(b)).
This regime is in the center of Fig.~\ref{DimerRegimes} and is limited to the left by the dashed dotted line and to the right by the dashed line. The dashed line corresponds to $\vartheta=\arccos(1/\sqrt{3})$. In the third region $V_1(x)>0$ for all $x$ and there is never a bound dimer. This situation occurs to the right of the dashed line in Fig.~\ref{DimerRegimes}. We note that for a single dimer it does not matter whether the molecules are fermions or bosons as long as tunneling between the layers can be neglected. Finally, the intertube interaction is only weakly dependent on the transverse confinement strength for $\lambda\gg1$ as the correction to the strict 1D limit scales as $\lambda^{-2}$.

\begin{figure}[t!]
\begin{center}
\includegraphics[width=0.8\linewidth,angle=0]{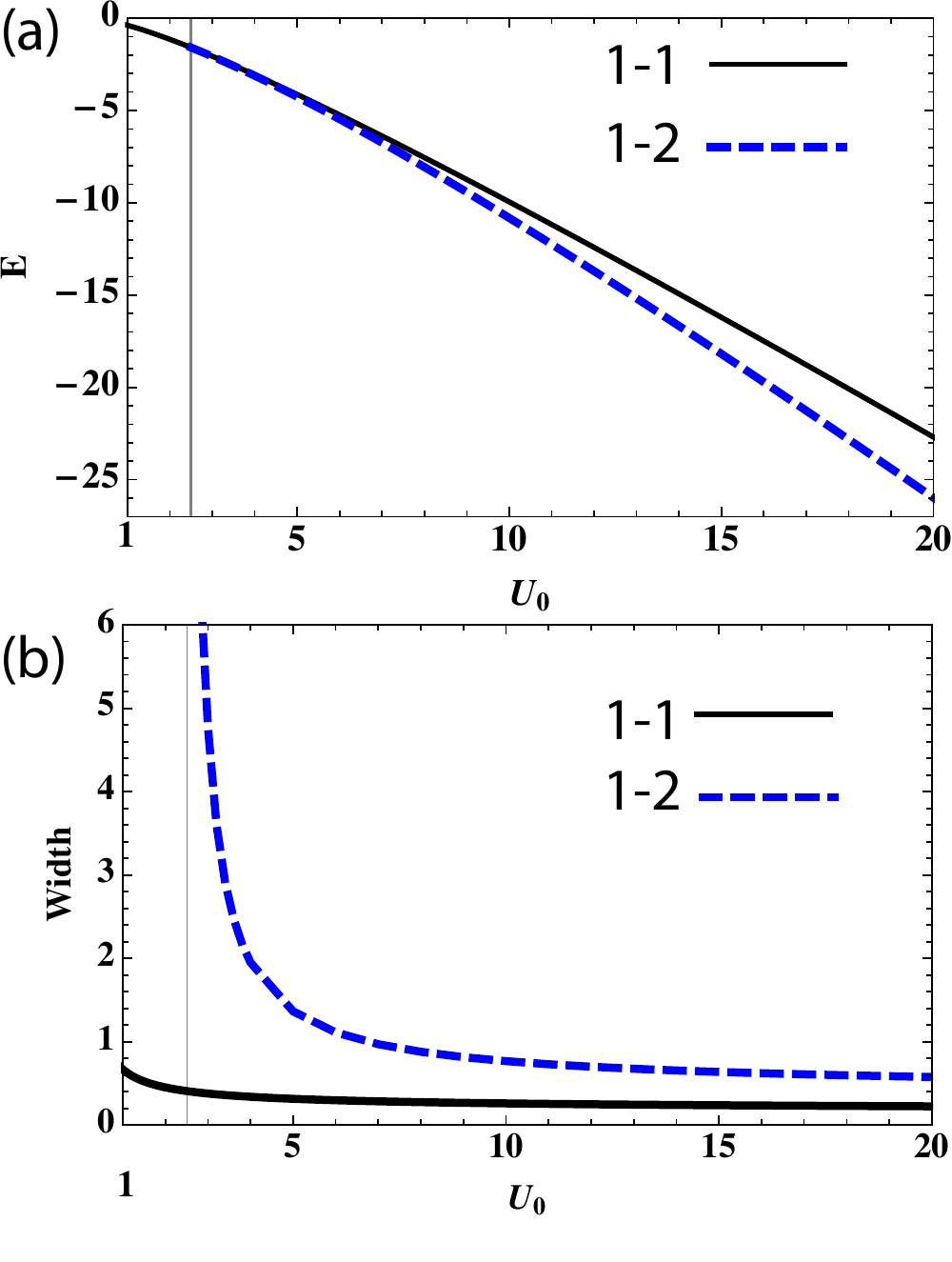}
\caption{(Color online) (a) Energy of the dimer state 1-1 (solid (black) line) and the trimer state 1-2 (dashed (blue) line) for $\varphi=57^\circ$ and $\lambda=10$. Vertical gridline at $U_0=2.5$  indicates critical interaction strength above which trimer is stable. For the parameters shown here bosonic and fermionic trimer have almost the same energy and larger few-body complexes are unstable. (b) Spatial extension of dimer and trimer as defined in Eq.~\eqref{eq:LengthTr}. Energy in units of $\hbar^2/m\Delta^2$ length in units of $\Delta$. }
\label{Fig:ETrphi57}
\end{center}
\end{figure}

\subsection{Trimer}
Next we discuss the trimer, which consists of one particle in one tube and two particles in the other tube as depicted in  Fig.~\ref{Fig:Complexes}(b).
As is the case of the dimer, the intertube attraction is responsible for binding the trimer. For a trimer to be stable, the energy gain from intertube attraction has to overcome the energy cost from intratube repulsion and from localization. In this subsection we consider $\vartheta=0$ where the intertube attraction is strongest.
The stability of the trimer can be increased by tilting the dipoles close to the magic angle $\varphi_{M}=\arccos(1/\sqrt{3})$ or by increasing $U_0$. The latter reduces the cost of localization energy and the former reduces the intratube repulsion. Figure~\ref{Fig:ETrphi57}(a) shows how the trimer becomes stable as the interaction strength $U_0$ is increased at a tilting angle $\varphi=57^\circ$ close to the magic angle.
Above a critical interaction strength of $U_0=2.5$ the energy of the trimer is smaller than three free molecules as well as one free molecule and one dimer.  Experimentally $U_0$ can be tuned by the strength of the homogeneous electric field. However, in order to reach large values of $U_0$, which are needed here, one needs to choose heteronuclear molecules with a large electric dipole e.g. $^6$Li$^{133}$Cs.

The instability of the trimer at small interaction strength $U_0$ is also visible in the spatial extension of the trimer.
Denoting by $x_n$ the position of particle $n$ and by $X_c=N^{-1} \sum_n x_n$ the center-of-mass, where $N$ is the total particle number,  we define the spatial extension of a  complex as
\begin{eqnarray}
l^2=\frac{1}{N} \sum_n (x_n-X_c)^2=\frac{1}{N^2} \sum_{m<n}(x_n-x_m)^2\label{eq:LengthTr}.
\end{eqnarray}
Fig.~\ref{Fig:ETrphi57}(b) shows the spatial extension of the dimer and trimer. The dissociation of the trimer into a free particle and a dimer at small $U_0$ leads to a diverging the spatial extension of the trimer for $U_0<2.5$. We note that the energy of the trimer smoothly approaches the energy of the dimer.

\begin{figure}[t!]
\begin{center}
\includegraphics[width=\linewidth,angle=0,scale=1]{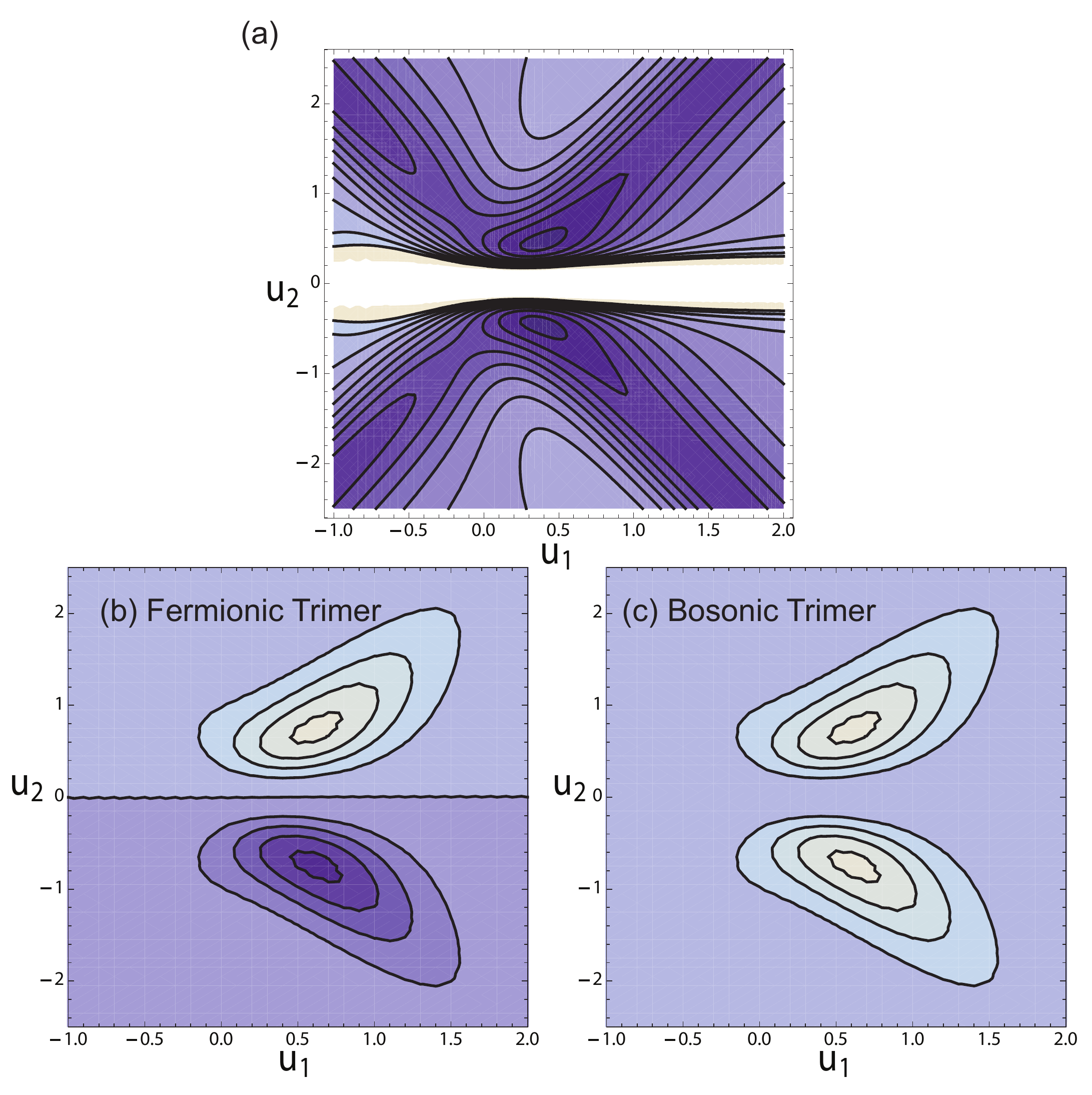}
\caption{(Color online) (a) Contourplot of the the trimer potential~\eqref{eq:trimer} for $\varphi=57^\circ$ as function of the relative coordinates $u_1=(-2 x_0+x_1+x_2)/\sqrt{6}$ and $u_2=(x_1-x_2)/\sqrt{2}$ in units of $\Delta$, see Eq.~\eqref{eq:relTr}. Intralayer repulsion gives a large potential barrier at $u_2=0$. Brighter areas correspond to higher values. (b), (c) Controuplot of the wave function for fermionic and bosonic trimers. The bosonic wave function is almost given by magnitude of the fermionic wave function which is expected for hard core bosons. Here $U_0=10$ while the remaining parameters are as in Fig.~\ref{Fig:ETrphi57}.}
\label{Fig:WFphi57}
\end{center}
\end{figure}

For the tilting angle  $\varphi=57^\circ$ and the interaction strengths shown in Fig.~\ref{Fig:ETrphi57}, the energy of the trimer is almost independent of whether the molecules are bosons or fermions.  The reason is the strong intratube repulsion at short distances, which turn bosons into hard core bosons that behave like fermions in 1D. This can be appreciated in Fig.~\ref{Fig:WFphi57}. The intratube repulsion causes a large potential barrier at $u_2=0$  as shown in Fig.~\ref{Fig:WFphi57}(a). As shown in  Fig.~\ref{Fig:WFphi57}(b) and (c), the wave function of the bosonic trimer is to a good approaximation given by the magnitude of the wave function of the fermionic trimer.

\begin{figure}[htb!]
\begin{center}
\includegraphics[width=\linewidth,angle=0,scale=1]{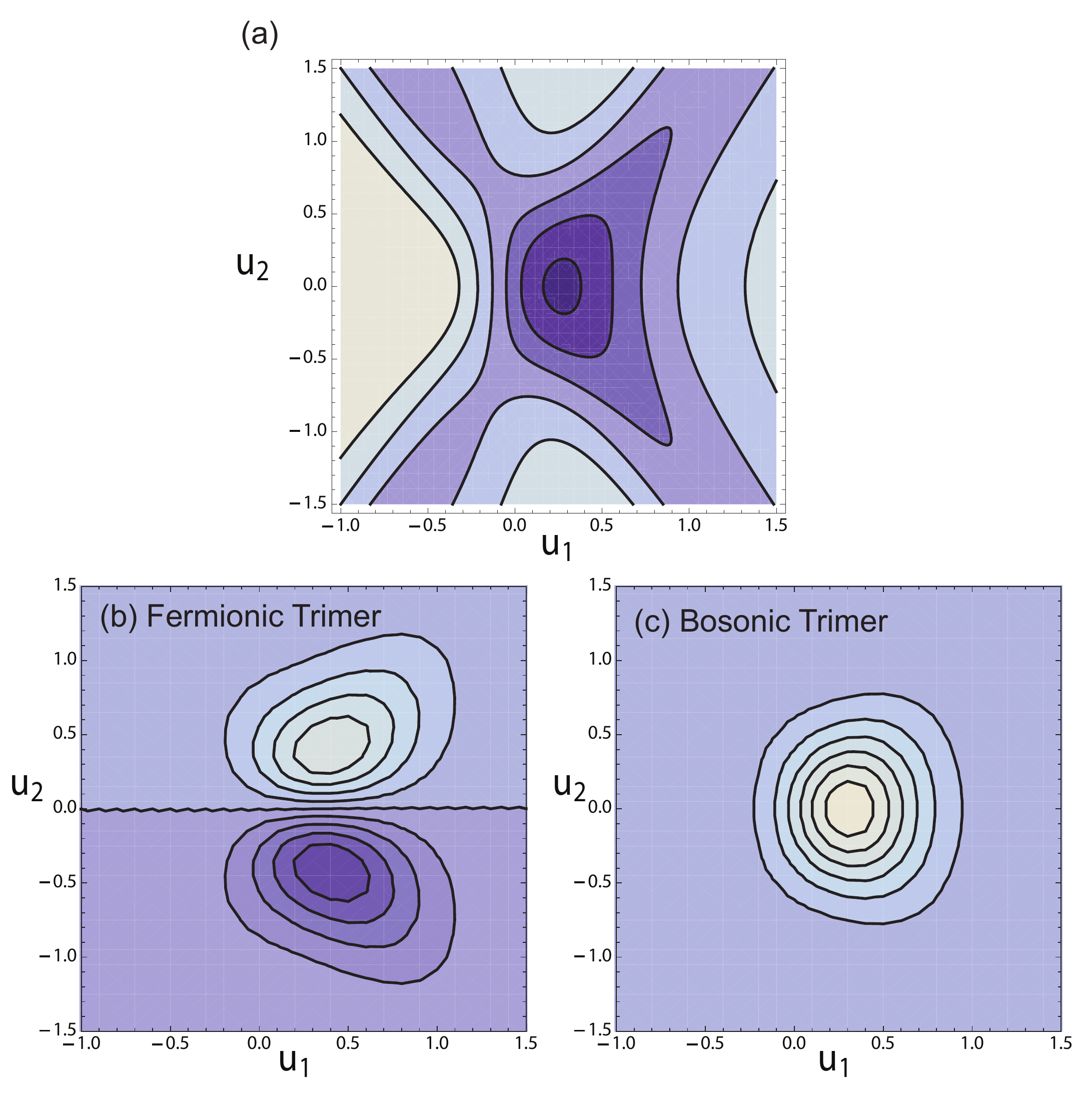}
\caption{(Color online) Same as Fig.~\ref{Fig:WFphi57} but for $\varphi=\varphi_M$, where intratube interaction vanishes. (a) Contourplot of the trimer potential shows a minimum at $u_2=0$. (b) Fermionic trimers has a node at $u_2=0$ due to Fermi-statistics. (c) Bosonic trimer has maximum at $u_2=0$. }
\label{Fig:WFc}
\end{center}
\end{figure}

At the magic angle, the effective intratube interaction vanishes and the difference between trimers made up of either bosonic or fermionic molecules is maximal.
Fig.~\ref{Fig:WFc}(a) shows the potential, $V_{1-2}$, of the trimer and the wave functions for bosonic or fermionic molecules. The relative wave function for the fermionic and bosonic trimer are shown in Fig.~\ref{Fig:WFc}(b) and (c).
For fermions, the antisymmetry of the wave function under particle exchange enforces that the wave function vanishes at $u_2=0$ although the potential is minimal there. Due to the antisymmetry of the wave function the fermionic trimer is only stable above a critical interaction strength even at the magic angle. In contrast, the wave function of the bosonic trimer has a maximum at $u_2=0$ for $\varphi=\varphi_M$ as shown in Fig.~\ref{Fig:WFc}(c) and the bosonic trimer forms at arbitrarily small $U_0$ at the magic angle.

\begin{figure}[t!]
\begin{center}
\includegraphics[width=0.8\linewidth,angle=0]{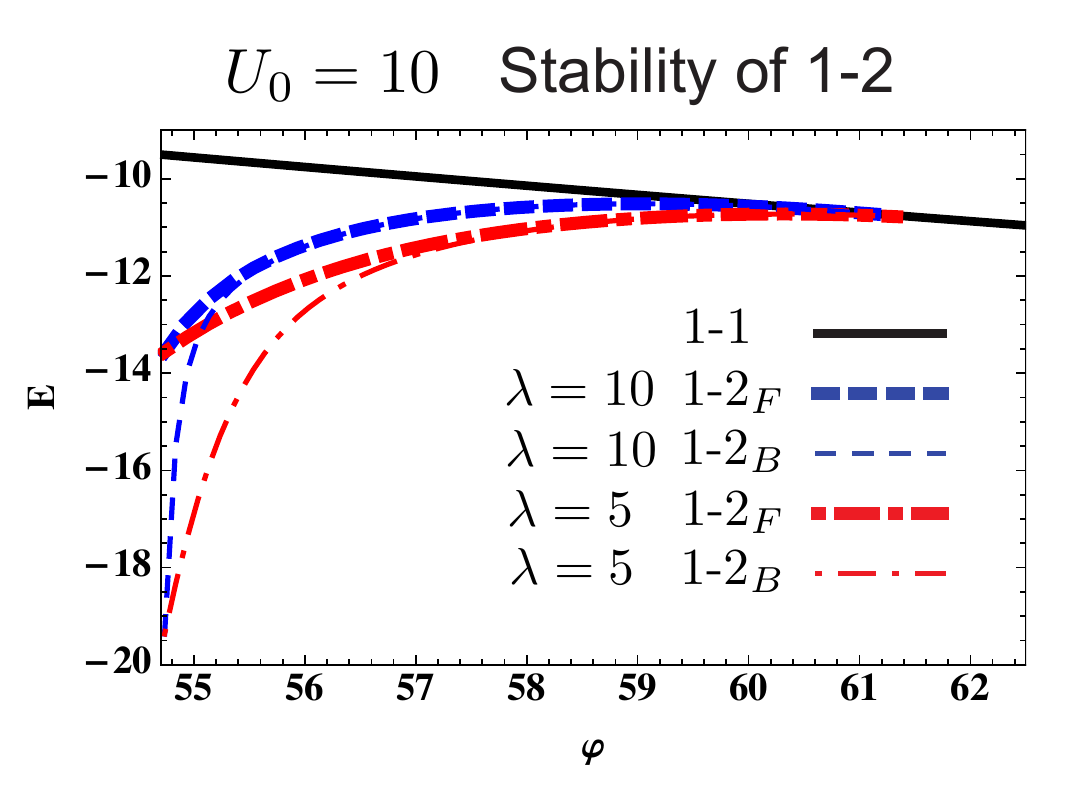}
\caption{(Color online) Energy of dimer and trimer few-body states as function of the tilting angle $\varphi\geq \varphi_M$ for fixed interaction strength $U_0=10$. The subscripts $F$ and $B$ indicate fermionic or bosonic molecules.
For strong lateral confinement ($\lambda\gg1$), the dimer energy is independent of particle statistics and on the lateral confinement. In contrast, the trimer energy depends on both $\lambda$ and statistics, in particular close to the magic angle $\varphi_{M}\approx 54.73^\circ$ where intratube interaction is strongly suppressed. At the magic angle only particles in different tubes interact with each other and this interaction is almost independent of $\lambda$. Note that we compare total energy of dimer and trimer although they consist of different number of particles. Energy in units of $\hbar^2/m\Delta^2$. }
\label{Fig:U10Tr}
\end{center}
\end{figure}

Fig.~\ref{Fig:U10Tr} shows the stability of the trimer states both for fermionic and bosonic molecules as function of the tilting angle $\varphi$ for  transverse confinement strengths $\lambda=5$ and 10.
The solid black line shows the energy of the dimer which has little dependence on the tilting angle for the regime shown. The thick dashed line shows the energy of the fermionic trimer at a strong transverse confinement of $\lambda=10$. At the magic angle its energy is minimal since intratube repulsion vanishes. By increasing $\varphi$, the intratube repulsion increases and at about $\varphi=60.5^\circ$ the trimer becomes unstable. The thick dash-dotted line shows the energy of the fermionic trimer at a smaller transverse confinement $\lambda=5$. Lowering the transverse confinement reduces the intratube repulsion at short length scales and thus lowers the energy of the trimer. At the magic angle, the intratube vanishes exactly and within our model (which assumes $\lambda\gg 1$) the energy of the trimer is independent of $\lambda$. For the fermionic trimer, the transverse confinement has a rather small effect, since the short length repulsion is less important due to the Fermi-statistics that forbids two fermions to be at the same point. However, as we already discussed in Fig.~\ref{Fig:WFc} bosonic molecules can share the same single-particle wave function and therefore bosonic trimers have a much lower energy than fermionic ones for small intratube repulsion. The energy of bosonic trimers is shown by the thin lines in  Fig.~\ref{Fig:U10Tr}. The energy of the bosonic trimers increases much faster than for fermionic trimers as $\varphi$ is moved away from the magic angle. Additionally, the transverse confinement has a stronger effect on the energy for bosonic molecules than for fermionic molecules. However, away from the magic tilting angle the intratube repulsion is sufficiently strong to suppress the overlap between the two molecules in the same tube and the bosonic and fermionic trimers have nearly the same energy. This situation is shown in Fig.~\ref{Fig:WFphi57}.

\subsection{Four-body states}
There are two possible four-body states, the 2-2 state with two particles in each tube and the 1-3 state with one particle in one tube and three particles in the other tube. They are shown schematically in Fig.~\ref{Fig:Complexes}(c) and (d). Similar to the trimer, the stability of these states is limited to tilting angles close to the magic angle $\varphi_M$ and sufficiently large interaction strength $U_0$. Again this is needed so that the attractive intertube interaction overcomes the repulsive intratube interaction and the cost of localization. For the parameters shown in Fig.~\ref{Fig:ETrphi57}, the four-body states are unstable. However, closer to the magic angle and for sufficiently strong interaction the four-body states are stable.

\begin{figure}[t!]
\begin{center}
\includegraphics[width=0.8\linewidth,angle=0]{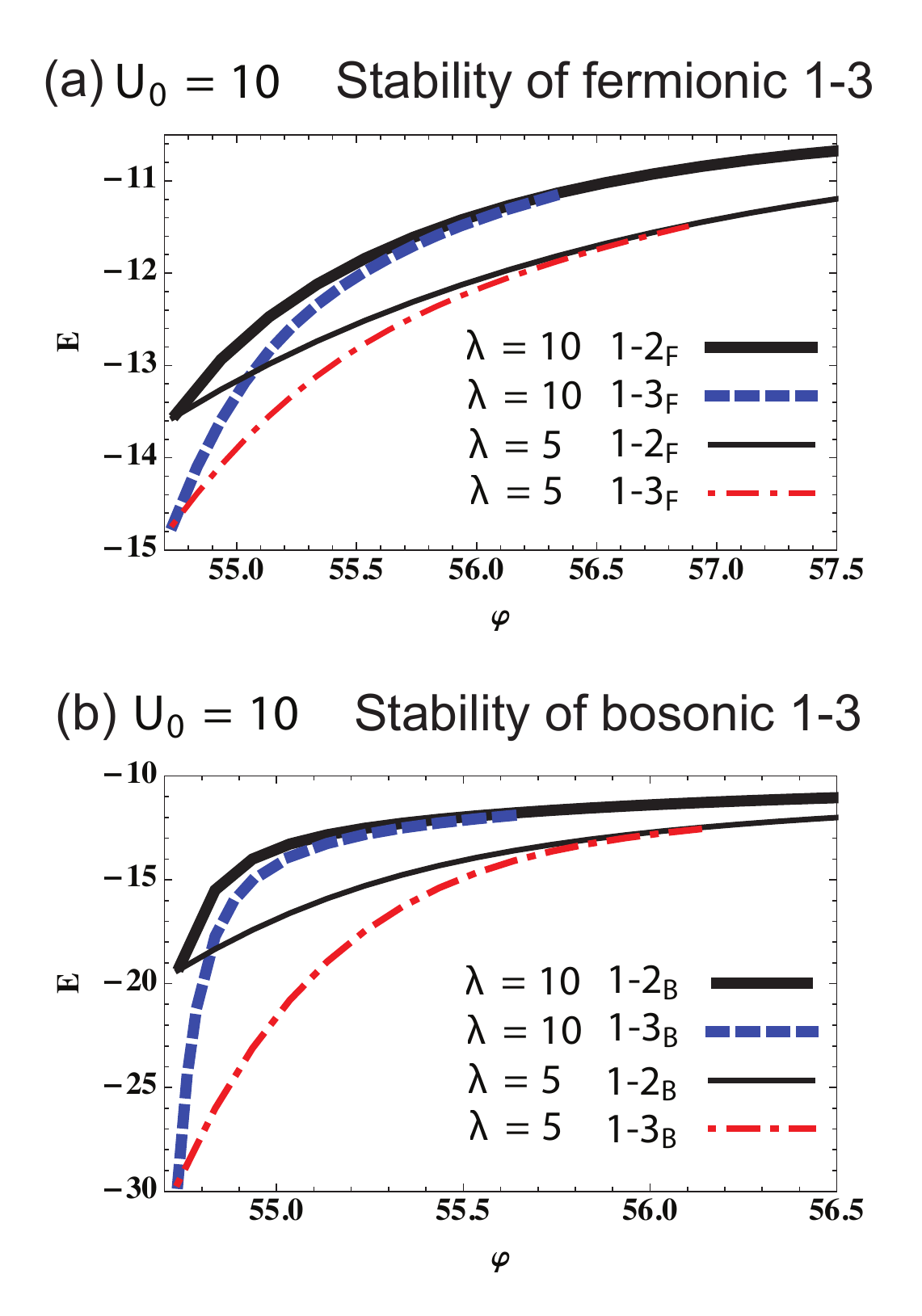}
\caption{(Color online) Energy of trimer and four-body states in the 1-3 configuration as function of the tilting angle $\varphi\geq\varphi_M$ for fixed interaction strength $U_0=10$.
Both 1-2 and 1-3 states depend on lateral confinement $\lambda$ and statistics. The stability of the 1-3 state is limited to a small interval of tilting angles close to the magic angle $\varphi_{M}\approx 54.73^\circ$. Again we note that we compare total energy of 1-2 and 1-3 complexes. Energy in units of $\hbar^2/m\Delta^2$. }
\label{Fig:U10F2}
\end{center}
\end{figure}

Fig.~\ref{Fig:U10F2} shows the stability of the 1-3 states both for fermionic and bosonic molecules as function of the tilting angle $\varphi$ and for different transverse confinement strengths $\lambda$.
Fig.~\ref{Fig:U10F2}(a) shows the fermionic case. The thick lines correspond to a transverse confinement $\lambda=10$. For $\varphi<56.3^\circ$ the 1-3 state (thick dashed line) is stable and has a lower energy than the 1-2 trimer (thick solid line). For $\varphi>56.3^\circ$ the 1-3 complex dissociates in a trimer and a free particle. The dissociation can also be seen in the spatial extension of the 1-3 complex (not shown here) which diverges for $\varphi>56.3^\circ$ in a similar way as for the trimer in Fig.~\ref{Fig:ETrphi57}.
The thin lines in Fig.~\ref{Fig:U10F2}(a) correspond to $\lambda=5$ and show qualitatively the same behavior.  Fig.~\ref{Fig:U10F2}(b) shows the bosonic case. We note that for bosons the complexes have a much lower energy than for fermions at the magic angle $\varphi_M$. However, the energy rapidly increases for larger tilting angles. For bosons, the short distance behavior of the intratube interaction is relevant close to the critical angle and there is a strong dependence on the transverse confinement.

\begin{figure}[t!]
\begin{center}
\includegraphics[width=0.8\linewidth,angle=0]{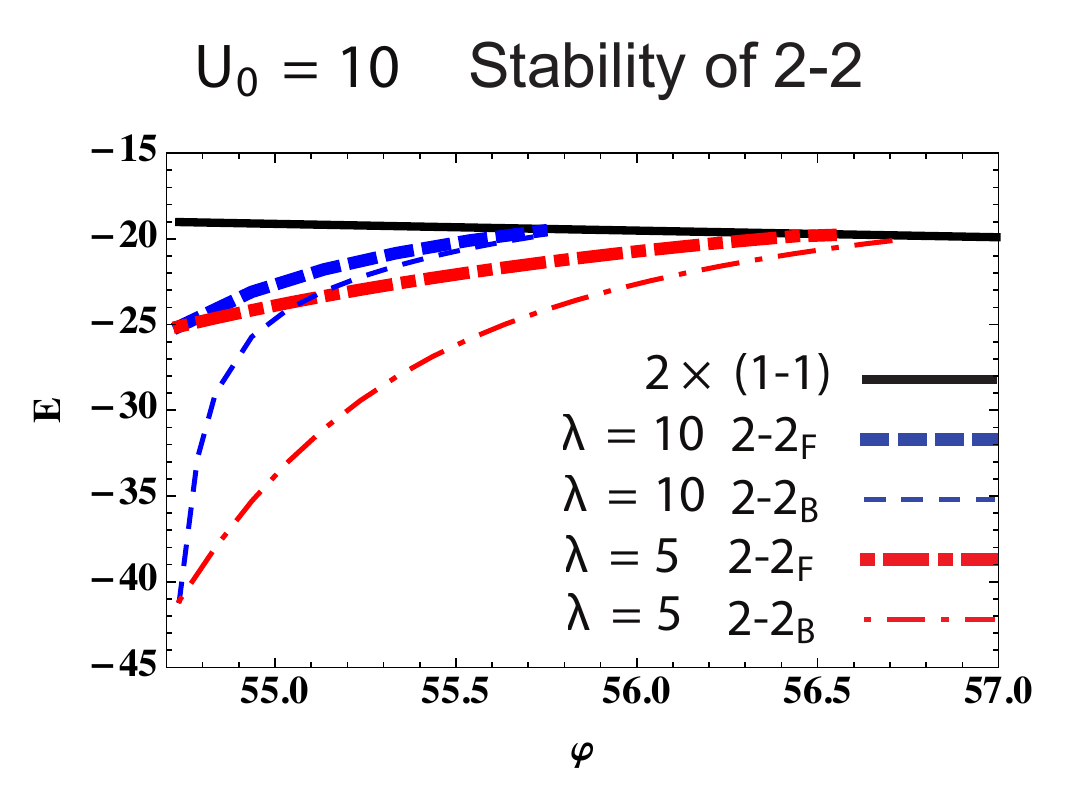}
\caption{(Color online) Energy of four-body states in the 2-2 configuration and the energy of two independent dimers as function of the tilting angle $\varphi\geq\varphi_M$ for fixed interaction strength $U_0=10$.
While the dimer energy is independent of $\lambda$ and statistics, the 2-2 state depends on both. We note that this time we compare complexes with the same number of particles. Energy in units of $\hbar^2/m\Delta^2$. }
\label{Fig:U10F1}
\end{center}
\end{figure}

Fig.~\ref{Fig:U10F1} shows the stability of the 2-2 states both for fermionic and bosonic molecules as function of the tilting angle $\varphi$ and for different transverse confinement strengths $\lambda$. At the magic angle the 2-2 state has lower energy than two independent dimers. Increasing intratube repulsion by increasing the tilting angle the 2-2 states soon become unstable. Notice again that bosonic states have a much smaller binding energy than fermionic ones.

\begin{figure}[t!]
\begin{center}
\includegraphics[width=0.95\linewidth,angle=0]{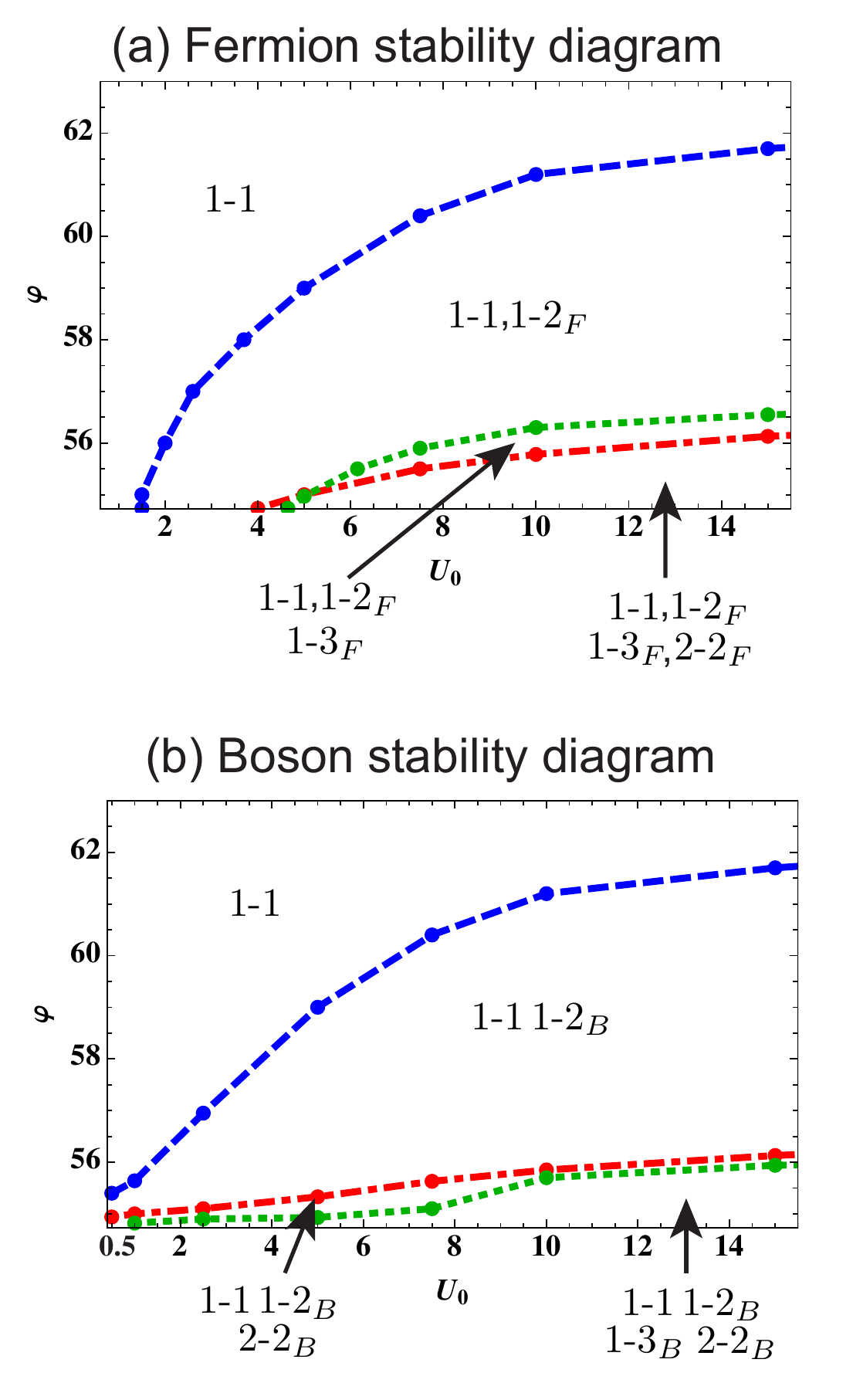}
\caption{(Color online) Stability of different complexes as function of interaction strength $U_0$ and tilting angle $\varphi$ for $\lambda=10$. Each region is labeled by its stable complexes. For large tilting angle, only the dimer is stable but as one approaches the magic angle $\varphi_M$, other complexes become stable. Below the (blue) dashed line the trimer is stable, below the (green) dotted line the 1-3 state is stable and below the (red) dashed-dotted line the 2-2 state is stable. (a) Fermionic molecules. All few-body states except the dimer have a critical interaction strength. (b) Bosonic molecules. Close to the magic angle the complexes become stable for any interaction strength.}
\label{Fig:ComplexesStability}
\end{center}
\end{figure}

\subsection{Stability diagram}
The regions of stability of the various complexes in the $U_0$-$\varphi$ plane is shown in Fig.~\ref{Fig:ComplexesStability}.  Above the (blue) dashed line only the dimer is stable, while below it the trimer also becomes stable. The (green) dotted line marks the position at which the 1-3 state becomes stable and below the (red) dashed-dotted line the 2-2 state is stable. The diagram shows that with increasing number of particles in the complex the stability regime is reduced to a small intervall starting at the magic angle. Furthermore, by comparing the stability diagram for fermionic and bosonic molecules a major difference is evident. For fermions, there is a critical interaction strength even in the absence of intratube repulsion, while bosons do not have a critical value.

\begin{figure}[t!]
\begin{center}
\includegraphics[width=0.8\linewidth,angle=0]{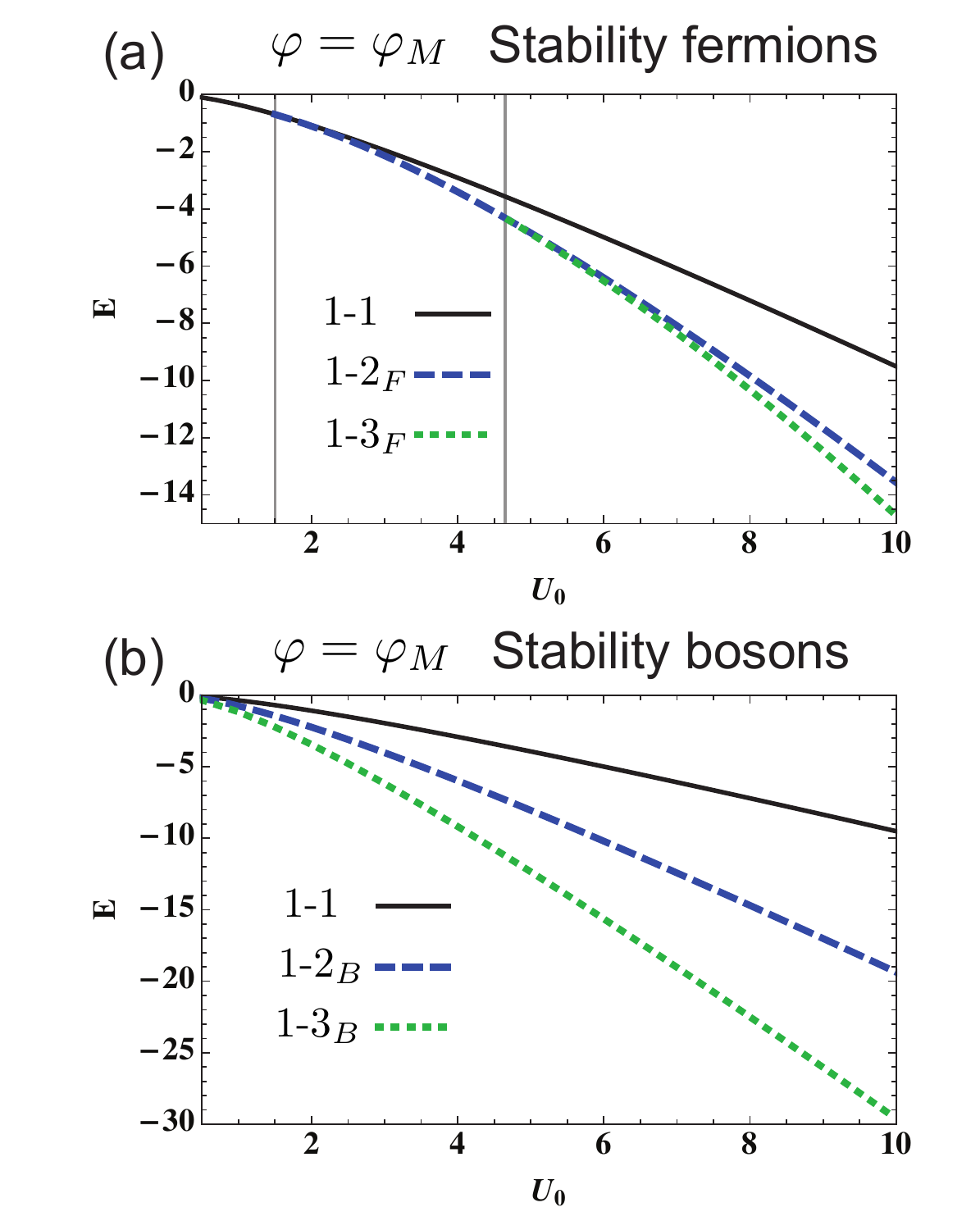}
\caption{(Color online) Energy of few-body states 1-1, 1-2 and 1-3 in units of $\tfrac{\hbar^2}{m\Delta^2}$ at the magical angle $\varphi_M$ as function of $U_0$ for $\lambda=10$. (a) Even in the absence of intratube repulsion the 1-2 and 1-3 configurations of fermonic molecules are only stable above critical interaction strength indicated by vertical lines. (b) For bosonic molcules, however, the 1-2 and 1-3 configurations are stable at any interaction strength at $\varphi=\varphi_M$. }
\label{Fig:PhicF2}
\end{center}
\end{figure}

\begin{figure}[t!]
\begin{center}
\includegraphics[width=0.8\linewidth,angle=0]{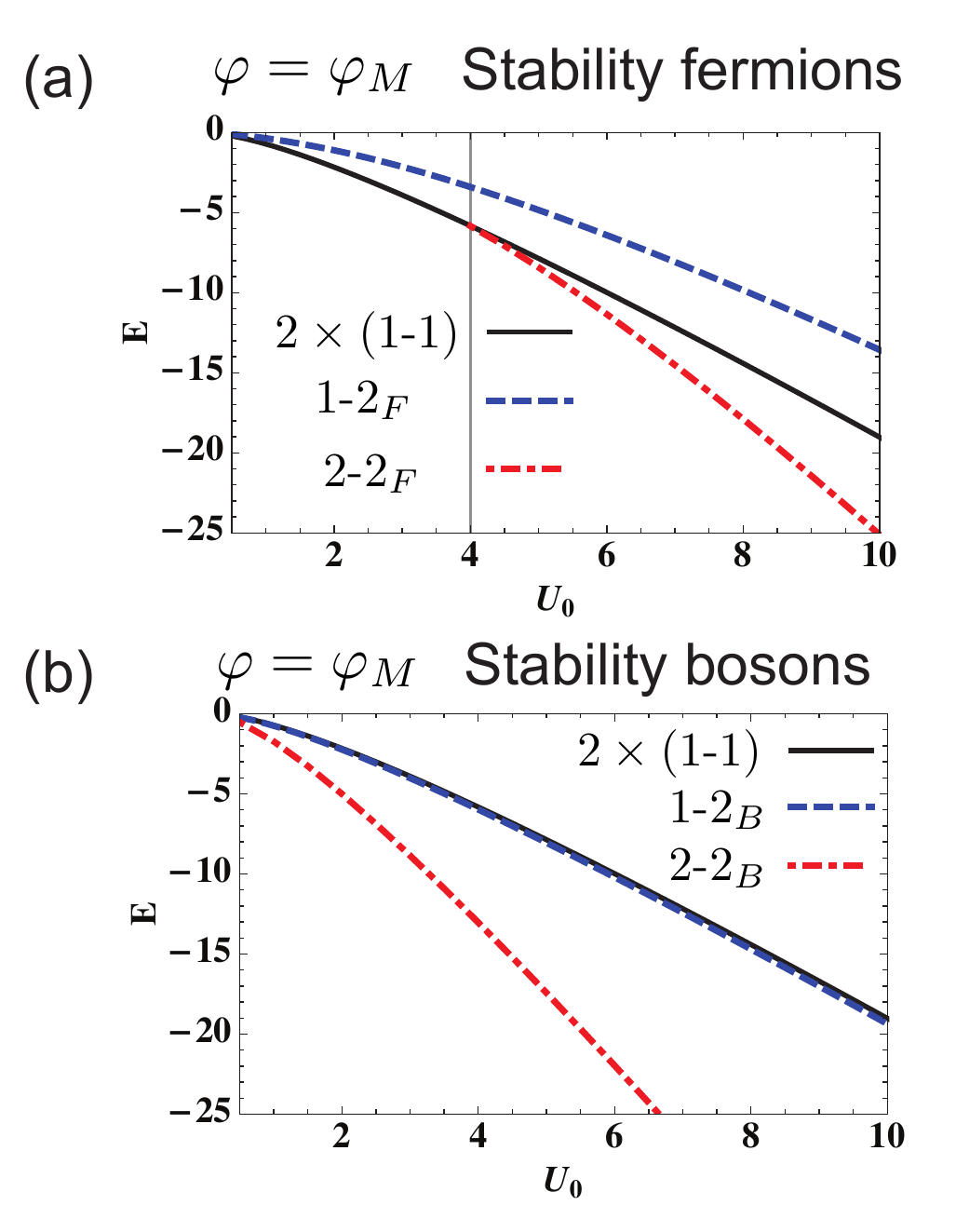}
\caption{(Color online) Energy of few-body states 1-1, 1-2, 2-2 in units of $\tfrac{\hbar^2}{m\Delta^2}$ at the magical angle $\varphi_M$ as function of $U_0$ for $\lambda=10$. (a) Even in the absence of intratube repulsion the 2-2  configurations of fermonic molecules is only stable above critical interaction strength indicated by vertical lines. (b) For bosonic molcules, however, the 2-2 configurations is stable at any interaction strength at $\varphi=\varphi_M$. Furthermore, the 1-2 configuration has even slightly lower energy than two independent 1-1 configurations. }
\label{Fig:PhicF1}
\end{center}
\end{figure}

Our results on the the stability of the few-body states at the critical angle are collected in Fig.~\ref{Fig:PhicF2} and \ref{Fig:PhicF1}.
As we can see complexes with  two or more fermionic molecules in the same tube are only stable above a critical interaction strength even in the absence of intratube interaction. Due to Fermi-statistics the wave function has to have a node for $x_1=x_2$, which increases the kinetic energy and increases the localization energy. In contrast, if the complex is made up of bosonic molecules, the wave functions can be finite at $x_1=x_2$ and there is no critical interaction strength. The regimes of stable trimers is extended for both bosons and fermions and lies in the range $\varphi_M\leq\varphi\leq 62^\circ$ for the values of $U_0$ shown in Figs.~\ref{Fig:PhicF2} and \ref{Fig:PhicF1}. This is a rather broad interval 
and implies that the study of trimers should be relatively easy. However, for the tetramer states the window of angles is much smaller and extends
only at most one or two degrees away from $\varphi_M$. Complexes larger than trimers therefore need a higher degree of fine-tuning of the setup and are more difficult to access.

\subsection{Larger Complexes}
For even larger complexes we could in principle continue the methods used here, calculate binding energies and
make the relevant comparisons. However, already from the data presented so far we can infer the influence of
such states. Consider the case of $N=5$ for which a 2-3 or a 1-4 configuration is possible. According to our arguments above, the stability regions for larger complexes in the bitube case will shrink to a small sliver just around $\varphi=\varphi_M$. In contrast, the three- and four-body
complexes are relevant over a broader range of parameter space, making them more accessible in
experiments.

\subsection{2D Layers}\label{sec2D}
When the tubes are replaced by 2D planes the extra non-confining dimension
increases the cost of localization and bound complexes are less stable.
However, we note that the classically preferred configurations are the same for bilayer and bitubes. Therefore trimers or even larger complexes will also be stable in bilayers, however, only for larger dipole strength \cite{volosniev2011b}.
An interesting difference between 1D and 2D concerns the dimer. The appearance of bound dimers in 2D is guaranteed at arbitrarily small couplings as long as the interlayer interaction integrates in space to zero or a negative value \cite{volosniev2011}. In 2D this precondition is fullfilled for any direction of the dipoles for the interaction \eqref{eq:dipole}.
In contrast, we discussed in Sec~\ref{sec:dimer} that in 1D the interlayer interaction can be made purely repulsive by increasing $\vartheta>54.7^\circ$. For purely repulsive interaction the dimer is of course unstable.

\section{Many-Body physics}\label{many-body}
At finite density one needs to minimize the free energy per particle. The stability of complexes is determined by comparing energies per complexes that can consist of different numbers of particles, e.g. trimer and dimer energies. If the interactions between complexes are neglected and the system is at equilibrium at a finite temperature, then stable complexes will be populated according to their energy per particle.
 At low temperatures and for balanced systems the occupation of trimers will be low, since the energy per particle is higher than in the dimer. However, for an imbalanced system the formation of larger complexes can be favorable even at low temperatures. For instance, if the number of particles in one tube is
twice that of the other, then the trimer would be favored over the dimer as soon as it is stable. For the
1-3 complex one would need a 1-3 ratio of the number of molecules in each tube and so forth.

For increasing densities interactions between complexes can no longer be neglected. An effective interaction between complexes can be obtained using the bound- state wave functions calculated here \cite{zinner2010}. Another important aspect is the statistics of the complexes \cite{santos2010}. While for bosonic molecules all complexes will behave as bosons, the character of complexes of fermionic molecules will change with the number of particles.  Then complexes with an odd number of molecules will have fermionic statistics, while even number complexes will behave as bosons.

In the dilute limit, we expect that when the binding energy of a
particular few-body state consisting of fermionic molecules
exceeds the Fermi energy then this state is well-defined. This situation will be realized in the strong-coupling limit where $U_0$ is
large. The Fermi energy depends on the density whereas $U_0$ depends on the molecules, the
electric field used to align them, and on the optical lattice needed to generate the
geometry. These two quantities can therefore be tuned independently.
From the strong-coupling side, the question then arises what the effective degree of freedom is. While the dimer and four-body can be considered as bosonic degrees of freedom, the
trimer is still a fermion. This means that systems dominated by one or the other
can have different quantum proporties.
By changing the direction or strength of the dipoles  we expect a crossover between fermionic and bosonic behavior, that  should be
observable using some of the techniques already proposed \cite{wang2006,wang2007,santos2010,potter2010}.

\section{Optical detection}\label{Sec:Detection}

The few-body complexes can be detected using light scattering. Recently, several nondestructive (in the sense of the quantum non-demolition, QND) schemes for measuring the properties of the many-body states in ultracold gases observing scattered light have been proposed \cite{PRL07,PRA07,LasPhys09,PRA09,Cirac08,Eugene09,Polzik09}. Among them, the method developed in Refs.~\cite{PRL07,PRA07,LasPhys09,PRA09} is the most relevant to the present system, as it explicitly uses the sensitivity of light scattering to the relative position of the particles forming a complex. This is due to the constructive or destructive interference of the light waves scattered from the different particles. Moreover, that method can be directly applied for extended periodic structures (many equidistantly spaced tubes or layers) and many-body systems, which makes the experimental realization promising.

We consider the scattering of the probe light with the amplitude given by the Rabi frequency $\Omega_p=d_0E_0/\hbar$ ($E_0$ is the probe-light electric field amplitude and $d_0$ is the induced dipole moment), cf. Fig.~\ref{Fig:Setup}. To increase the signal, the scattered light can be collected by a cavity, and the photons leaking from the cavity are then measured. Alternatively, the measurement of photons scattered can be made in a far-field region without the use of a cavity.

After several assumptions (the small tube radius, far off-resonant light scattering, detection in the far field zone), the light scattering has a simple physical interpretation. The scattered light amplitude is given by the sum of the light amplitudes, scattered from each molecule (cf. Fig.~\ref{Fig:Setup}, where several molecules are schematically depicted within two tubes). Each term has a phase and amplitude coefficient depending on the position of the molecule as well as on the  direction and amplitude of the incoming and outgoing light waves:
\begin{eqnarray}\label{1.5}
a_s=C\sum_{i=A,B}\int dx \hat{n}_i(x) u^*_s(x,\rho_i)u_p(x,\rho_i),
\end{eqnarray}
where the sum is over two tubes A and B, $\hat{n}_i(x)$ is the operator of particle linear density. For the free space scattering, the value of $C$ corresponds to the Rayleigh scattering \cite{Scully}. Adding a cavity to the setup the scattering is increased and $C=-ig_s\Omega_p/(\Delta_a\kappa)$ with $\kappa$ being the cavity decay rate, $g_s$ is the molecule-light coupling constant, and $\Delta_a$ is the light detuning from the resonance, cf. Refs.~\cite{PRL07,PRA07,LasPhys09,PRA09}.  In Eq.~(\ref{1.5}), $u_{p,s}(x,\rho_i)$ are the mode functions of probe and scattered light at the tube positions $\rho_{A,B}$, which contain the information about the propagation directions of probe and scattered light waves with respect to the tube direction. For the simplest case of two traveling light waves, the product of two mode functions takes the well-known form from classical light scattering theory: $u^*_s(x,\rho_i)u_p(x,\rho_i)=\exp{[i({\bf k}_p-{\bf k}_s){\bf r}_i}]$, where ${\bf k}_{p,s}$ are the probe and scattered light wave vectors, and ${\bf r}_i$ is the molecule position.

Equation~(\ref{1.5}) is valid for any optical geometry and can describe the angular distribution of the scattered light. However, an important conclusion of Refs.~\cite{PRL07,PRA07,LasPhys09,PRA09} was that some information about the many-body state can be obtained even by a simple measurement of the photon number scattered at a single particular angle, which is enough for our purpose. Moreover, as it was shown, the particularly convenient angle of measurement corresponds to the direction of a diffraction minimum, rather than Bragg angle (diffraction maximum). At the directions of diffraction minimum any classical (possibly very strong) scattering (leading to a signal scaling )is suppressed, and the light signal exclusively reflects the quantum fluctuations of the particles. 

We now fix the optical geometry as follows (cf. Fig. 1). The incoming probe light is a traveling or standing wave propagating at the direction perpendicular to the tubes, which gives $u_{p}({\bf r})=R(x)\exp(ik_py)$ (for the traveling wave) or $u_{p}({\bf r})=R(x)\cos(k_py)$ (for the standing wave) and includes the transverse probe profile $R(x)$ of an effective width $W$. To perform the measurements at the direction of a diffraction minimum, the scattered light is measured along $z$ direction. For the free space detection, or the traveling-wave cavity, this gives $u_{s}({\bf r})=\exp(ik_sz)$, while for the case of a standing wave cavity, $u_{s}({\bf r})=\cos(k_sz)$. Without loss of generality, we can assume $u_{s}({\bf r})=1$ at the tube position $z=0$. The absolute values of the wave vectors are equal to their vacuum quantities $k_{p,s}=2\pi/\lambda_\text{light}$.

An important property of such a configuration (illumination and detection at directions perpendicular to the tubes), is that all atoms within two different tubes scatter light with a fixed phase difference with respect to each other, independently of their longitudinal position $x$ within the tube. Here we assume a tight transverse confinement such that the motion of atoms in the perpendicular directions is frozen.
Due to this fact, the averaging over the probabilistic position of the complex does not involve the light phase. Moreover, averaging over the probabilistic relative positions within each complex does not involve the dependence on the light phase as well. At other directions, both those kinds of phase averaging are important and would decrease the optical signal and the distinguishability of the complex types. Our scheme is 
very convenient since detection takes place at a single angle perpendicular to the tubes only and requires no scan over multiple angles.

The operator of the light amplitude reduces to
\begin{eqnarray}\label{1.6}
a_s=C\left(u_p(y_A)\hat{N}_A(W)+u_p(y_B)\hat{N}_B(W)\right),
\end{eqnarray}
where $\hat{N}_{A,B}(W)$ are the operators of the effective particle numbers in the tubes A and B within the region illuminated by the laser beam,
\begin{eqnarray}\label{1.7a}
\hat{N}_{A,B}(W)=\int_{-\infty}^{\infty}\hat{n}_{A,B}(x)R(x)dx.
\end{eqnarray}
If the laser profile can be approximated by a constant in the interval $(-W/2, W/2)$, the operators $\hat{N}_{A,B}(W)$ exactly correspond to the atom number operators in two tubes within the laser beam.

The classical condition of the diffraction minimum is fulfilled, when the expectation value of the light-amplitude operator (\ref{1.6}) is zero due to the perfect cancelation of the expectation values of two terms in Eq.~(\ref{1.6}) (i.e. the total destructive interference between the scatterers in two tubes). This is achieved for $u_p(y_B)/u_p(y_A)=-\langle\hat{N}_A\rangle/\langle\hat{N}_B\rangle$. We introduce the atom number ratio $\alpha=\langle\hat{N}_A\rangle/\langle\hat{N}_B\rangle$. For the equal mean atom numbers (the few-body complexes 1-1 and 2-2), the optical geometry should be chosen such that $u_s(y_B)/u_s(y_A)=-1$, which can be achieved if, e.g., the tube spacing is the half of the light wavelength, $\Delta=\lambda_\text{light}/2$. For the few-body complex 1-2, $\alpha=1/2$, and the diffraction minimum is achieved if the light wavelength and tube spacing satisfy the condition $\cos(k_py_B)/\cos(k_py_A)=-1/2$. This can be achieved, e.g., if the position of the tube A corresponds to the antinode of the standing wave $\cos(k_py_A)=1$, while that of tube B corresponds to $k_py_B=2\pi/3$ or $4\pi/3$, leading to the ratios between the tube spacing and light wavelength as $\Delta=\lambda_\text{light}/3$ or $2\lambda_\text{light}/3$. Similarly, for the 1-3 complex, that ratio can be $\Delta\approx 0.3\lambda_\text{light}$ or $0.7\lambda_\text{light}$. All those example ratios can be indeed larger, taking into account the periodicity of the light wave.

The expectation value of number of photons scattered at the direction of diffraction minimum $n_{\Phi}$ is then given by
\begin{eqnarray}\label{1.7}
&n_{\Phi}=\langle a^\dag_s a_s\rangle =\left|C\right|^2 \left|u_p(y_A)\right|^2 \times&\nonumber\\
&\left\langle\left(\hat{N}_A(W)-\alpha\hat{N}_B(W)\right)^2\right\rangle,&
\end{eqnarray}
where $u_p(y_A)$ can be easily chosen as 1. This expression manifests that the number of photons scattered in the diffraction minimum is proportional to the second moment of the "rated" particle number difference between two tubes in the laser-illuminated region. In general, the photon number at the diffraction minimum is non-zero. It directly reflects the particle number fluctuations and correlations between the tubes. Thus, the number of photons reflects the quantum state of ultracold molecules.

\begin{figure}[htb!]
\begin{center}
\includegraphics[width=0.95\linewidth,angle=0]{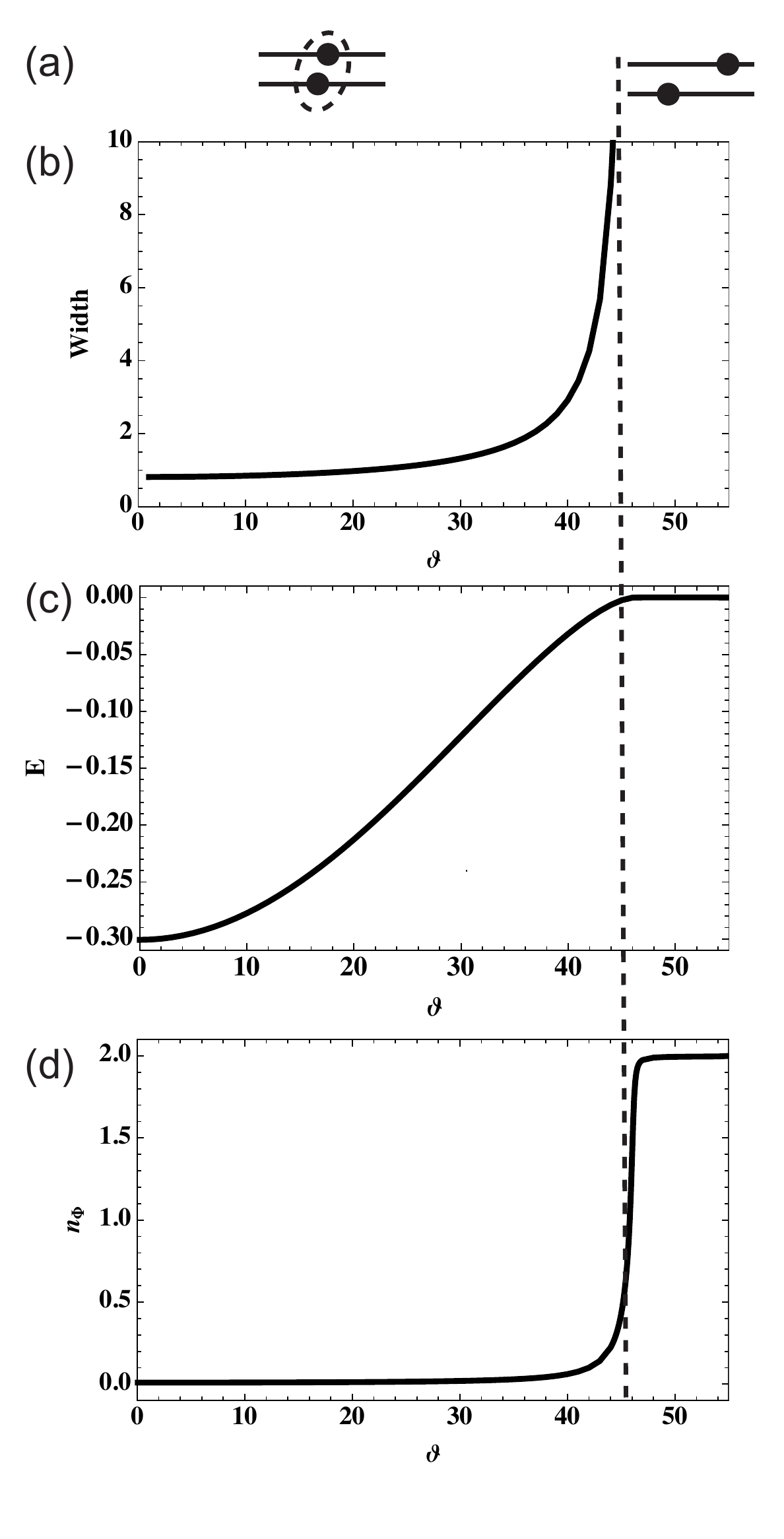}
\caption{(Color online) Optical detection of a dimer for $U_0=1/2$ and $\varphi=90^\circ$. (a) Visualization of two different regimes depending on the angle $\vartheta$. (b) Width of the dimer given by Eq.\ref{eq:LengthTr} in units of $\Delta$ between the dissociating particles. (c) Dimer energy in units of $\tfrac{\hbar^2}{m\Delta^2}$. (d) Number of photons scattered into diffraction minimum, Eq.\ref{IntDimer}, in unit of $n_0$: Two different photon numbers (zero and $2n_0$) correspond to the dimer and two free particles. The light intensity jump corresponds to the dissociation (or creation) of a dimer.}
\label{Fig:DetectionD}
\end{center}
\end{figure}

We now apply Eq.~\eqref{1.7} to the few-body states discussed in this paper. The calculations of the expectation values in Eq.~\eqref{1.7} simplifies strongly if one assumes the Gaussian shape of the probe light as $R(x)=\exp{(-x^2/W^2)}$ with the beam width $W$.

Applying Eq.~\eqref{1.7} to the case of one molecule per tube, where the formation of dimers is possible, we obtain the photon number $n_{\Phi}^{D}$ as
\begin{eqnarray}
n_{\Phi}^{D}&=&n_0 \left(1+\alpha^2-2\alpha \left \langle e^{-\frac{(x_0-x_1)^2}{2W^2}}\right \rangle \right),\label{IntDimer}
\end{eqnarray}
where $n_0=\left|C\right|^2 \left|u_p(y_A)\right|^2 (W/L)\sqrt{\pi/2}$ is an effective photon number, which can be scattered from a single molecule and takes into account the probability of finding a molecule in an illuminated fraction $W/L$ of the tube of the length $L$. The brackets denote the average over the relative wave function. As discussed in Appendix~\ref{RelCoord} one separates the center of mass and writes the two particle wave function as $\Psi(x_0,x_1)=\Psi_C(u_0) \psi(u_1)$ were $u_0,u_1$ are defined in Eq.\eqref{eq:relD} then 
\begin{align}
\left \langle e^{-\frac{(x_0-x_1)^2}{2W^2}}\right \rangle=\int\,du_1 e^{-\frac{u_1^2}{W^2}} \psi^2
(u_1).
\end{align}
Equation \eqref{IntDimer} describes the dependence of the photon number on the effective size of the dimer $x_0-x_1$. For a tightly bound dimer, the expectation value in Eq.~\eqref{IntDimer} is 
\begin{align}
\left \langle e^{-\frac{(x_0-x_1)^2}{2W^2}}\right \rangle=1, 
\end{align}
while for two independent molecules it approaches 0. Correspondingly, in a diffraction minimum ($\alpha=1$), the light scattering is completely suppressed for a dimer and approaches a constant $2n_0$ for two unbound molecules. This manifests the fact that for a dimer, the molecules appear in the laser beam only in pairs and the light scattered from the two molecules destructively interfere. In contrast, for the free molecules, there is a probability that only one molecule will be found in a laser beam, while another molecule is out of it. In the latter case, the perfect destructive interference is not possible, which leads to the nonzero photon number scattered in the diffraction minimum.

Figure 14 shows the striking sensitivity of the photon number to the presence of a dimer or free molecules. When the dimer dissociates during tilting of the $\vartheta$ angle, the photon number sharply jumps from the zero value to a value $2n_0$ (Fig. 14d). Figure 14 displays the dimer size and energy as well as the photon number. The latter shows a characteristic jump as the dimer becomes unstable above a critical angle $\vartheta$.

\begin{figure}[htb!]
\begin{center}
\includegraphics[width=0.8\linewidth,angle=0]{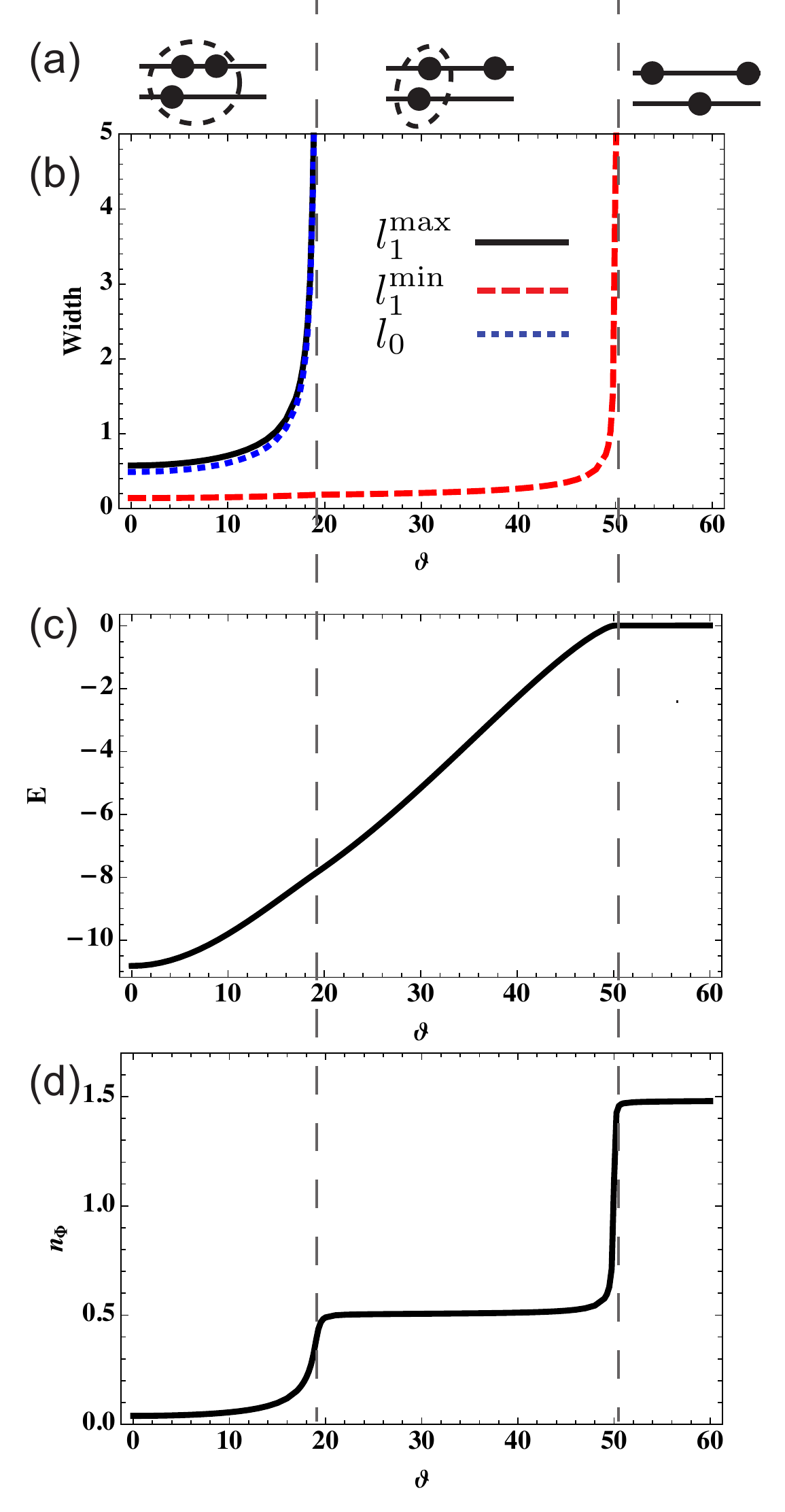}
\caption{(Color online) Optical detection of trimer for $U_0=10$ and $\varphi=57^\circ$. (a) Visualization of  three different regimes depending on the angle $\vartheta$. (b) Relative distances in units of $\Delta$ between the particles as defined in the text after Eq.~\eqref{eq:IntTr} showing the consecutive dissociation of the trimer into a dimer and unbound molecules. (c) Energy of the state in units of $\tfrac{\hbar^2}{m\Delta^2}$. (d) Number of photons scattered into diffraction minimum, Eq.~\eqref{eq:IntTr}, in units of $n_0$: Three different photon numbers (zero, $1/2n_0$, and $3/2n_0$) correspond to the trimer, dimer and a free molecule, and three free particles. The light intensity jumps correspond to the dissociations (or creations) of a dimer and trimer.}
\label{Fig:DetectionT}
\end{center}
\end{figure}

For the trimer case 1-2, the photon number reads
\begin{eqnarray}
\frac{n_{\Phi}^{Tr}}{n_0}&=& 1+2\alpha^2 +2\alpha^2 \left \langle e^{-\frac{(x_2-x_1)^2}{2W^2}} \right \rangle \notag\\
&&-2\alpha \left \langle e^{-\frac{(x_0-x_1)^2}{2W^2}} \right \rangle-2\alpha \left \langle e^{-\frac{(x_0-x_2)^2}{2W^2}} \right \rangle. \label{eq:IntTr}
\end{eqnarray}
As in this case, the population of the tubes is imbalanced, the diffraction minimum corresponds to $\alpha=1/2$ and for the tightly bound trimer, the photon number in the diffraction minimum is zero.
As previously, various terms correspond to the contribution of various relative coordinates between the molecules. For the tightly bound complex, all distances are small and the expectation values 
$\langle \exp(-(x_i-x_j)^2/2W^2)\rangle\approx 1$, while for large interparticle distances they vanish.

We are interested in three different regimes: 1) Trimer is stable. 2) Trimer is unstable but dimer is stable. 3) Trimer and dimer are unstable.
In order to distinguish these regimes we do not only calculate the extension of the trimer $l$, but calculate the following quantities 
(where the coordinates are given in Fig~\ref{Fig:Complexes})
\begin{eqnarray}
l^2&=& (l_1^{\text{max}})^2+(l_1^{\text{min}})^2+l_0^2,\\ 
(l_1^{\text{max}})^2&=&\left\langle \text{Max}[(x_0-x_1)^2,(x_0-x_2)^2]/9\right\rangle, \\
(l_1^{\text{min}})^2&=&\left\langle \text{Min}[(x_0-x_1)^2,(x_0-x_2)^2]/9\right\rangle, \\
l_0^2&=&\left\langle (x_1-x_2)^2/9\right\rangle.
\end{eqnarray}
Here $l_0$ is the intratube distance measuring the average distance between the two molecules in the same tube.
There are two intertube distances, $l_1^{min},l_1^{max}$, between the single molecule in tube $A$ and the two molecules in tube $B$.
$l_1^{min}$ denotes the smaller and $l_1^{max}$ the larger intertube distance.

Figure 15 shows the relative sizes, energy and photon number for the three-body complexes.
At small $\vartheta$ the trimer is stable, all length are finite and the intensity of the scattered light is approximately zero.
At $\vartheta=19^\circ$ the trimer dissociates in the dimer and a free molecule.
Then the larger of the intertube distances $l_1^{\text{max}}$ diverges and the light intensity jumps up, because the condition of the perfect destructive interference is not satisfied (the free molecule can be within or outside the probe beam). Interestingly, the energy of the trimer smoothly approaches the energy of the dimer and therefore the energy shows no indication of the dissociation. In contrast the scattered light intensity jumps to a new plateau.
At even larger angles At $\vartheta>51^\circ$ the dimer also dissociates and the system consists of three free molecules.
Now all distances diverge and the light intensity increases further as well, manifesting the increased molecule number fluctuations in the laser beam waist. As shown in Fig. 15, all three possible molecular states can be well distinguished by light scattering and correspond to three different light intensities: zero for a trimer, $n_0/2$ for a dimer and a free molecule, and $3 n_0/2$ for three free molecules. The complex dissociation corresponds to the jump in the photon number. When $\vartheta$ decreases, the creation of the complexes is manifested by the stepwise decrease in the light intensity, which corresponds to the reduced atom particle fluctuations.

Notice that our detection scheme uses the difference between plateaus to detect 
the few-body complexes. In a real experiment, there could be additional 
background noise added to the signal. However, the difference in the 
height of the plateaus stays the same. This relative measurement of 
heights is therefore more sensitive than an absolute one, and can 
help eliminate problems from constant noise in experiments.

In the many-body context, when the density of molecules is small, the results for 
the number of scattered photons Eqs.~\eqref{IntDimer} and \eqref{eq:IntTr} 
(displayed in Figs.~\ref{Fig:DetectionD}(d) and \ref{Fig:DetectionT}(d)).
still hold and should be multiplied by the number of molecules. The advantage 
of using the diffraction minumum is that the classical coherent scattering
which scales with the number of molecules squared is suppressed, and we 
get our signal from the fluctuations which scale with the number of 
molecules. We still require a low density in order to allow the molecules
to dissociate to larger distances. In a high density limit, it would not 
make sense to go beyond the average particle distance. When this becomes
comparable to the complexes size, we have effectively a crystal configuration
and no jumps in the signal.

The analysis of the four-body complex 1-3 can be carried out in a similar way. 
It results in three steps in the photon number dependence on $\vartheta$, which 
correspond to the 1-3 complex, 1-2 trimer and 1 free molecule, 1-1 dimer and 2 
molecules, and four unbound molecules.

While the photon signal from single molecules can be small, we have outlined 
a number of ways above to enhance the signal. This can be done by increasing 
the number of molecules in the tubes, increasing the probe laser intensity,
the measurement time, or by using a high-quality cavity. Additionally, we 
note that an array of many tubes in one plane would also enhance the signal 
proportionally to the number of tubes due to the perpendicular arrangement 
of probe and detection. Which one of these mechanisms is the better option
depends on concrete experimental setups.

\section{Experimental Issues}
In order to see the complexes discussed here in an experiment it is the value of $U_0$ that matters.
Our calculations indicate that the trimer should be stable from $U_0>1.5$ at the magic angle $\varphi=\varphi_M$ 
(see Fig.~\ref{Fig:PhicF2}), whereas for a larger angle of $\varphi=57^\circ$ it is stable for $U_0>2.5$ (see Fig.~\ref{Fig:ETrphi57}).
A minumum dipoles strength is therefore required to see complexes beyond the dimer in current experiments with fermionic polar
molecules.
The strength parameter can be written in
the convenient form
\begin{align}
U_0=0.15\left(\frac{m}{1\,\text{amu}}\right)\left(\frac{1\,\mu}{\Delta}\right)\left(\frac{D}{1\,\text{Debye}}\right)^2,
\end{align}
where 1 Debye is $3.336\cdot 10^{-30}$ Cm in SI units. For current experiments
using $^{40}$K$^{87}$Rb the 
maximum dipole moment is 0.56 Debye \cite{ospelkaus2008,ni2008,ospelkaus2010,ni2010}. 
Assuming $\Delta=532$ nm we obtain $U_0=1.12$ which is a rather small value for investigating
few-body complexes. Working with a different molecule of larger moment therefore seems appealing.
For instance, a molecule like $^6$Li$^{133}$Cs should be able to accomodate much larger dipole
moments \cite{deiglmayr2008} and by tuning the external field a large range of $U_0$ should be accessible
with that combination of alkali atoms. Another advantage of using a much larger dipole moment
is that the binding energy increases rapidly with $U_0$. This implies that one does not need to 
be in the degenerate regime and can study the bound complexes even with thermal samples.

An interesting alternative idea is to apply an optical lattice along the tubes in order to 
quench the kinetic energy of the molecules. This can be understood as increasing the 
effective mass, $m^*$, of the molecules over the bare value, $m^*/m>1$. This results
in a new effective dipole strength, $U^*=m^*U/m>U$. However, the introduction of 
such a lattice changes the dispersion relation of the molecules in the tubes. The 
study of bound states in periodic potentials with short-range interactions have demonstrated
modifications of the spectrum as the lattice depth is increased \cite{orso2005,wouters2006,valiente2010}.
Similar effects are likely for polar molecules when a periodic lattice potential is included in the 
bound state calculations. Such an elaborate extension is, however, beyond the scope of the 
present work and we include only an estimate of the influence of a lattice.

In order to study the fermionic trimer in the $^{40}$K$^{87}$Rb system we need an effective 
dipole strength of at least 1.5, or even larger to study the dependence on $\varphi$. 
For the sake of argument, let us assume that we are interested in obtaining
an effective $U^*\gtrsim 2U_0$. If we assume a tight-binding approximation, the effective
mass will then have to be $m^*\gtrsim 2m$. Consider a standard optical lattice potential of 
the form $V(x)=s E_0\sin^2(2\pi x/\lambda)$, where $E_0=2\hbar^2\pi^2/m\lambda^2$ is the 
recoil energy. Numerical calculations \cite{kramer2002} show that one needs $s=5$ to
have $m^*\sim 2m$ when two-body interaction are ignored. The tight-binding
approximation compares very well to a full numerical diagonalization for $s\geq 5$ \cite{kramer2002}.

The effect of interactions on the dispersion in the lattice is small unless the 
interaction energy considered is close to $E_0$. Let us for simplicity assume that 
the in-tube lattice has the same wavelength as the transverse lattice, i.e. $\Delta=\lambda/2$.
We then have $E_0=\tfrac{\pi^2}{2}\tfrac{\hbar^2}{m\Delta^2}\sim 4.9\tfrac{\hbar^2}{m\Delta^2}$.
From Fig.~\ref{Fig:ETrphi57} we see that the dimer and trimer binding energies for $U_0\sim 2.5$
is $\sim 2\tfrac{\hbar^2}{m\Delta^2}$. This implies that the interaction energy is smaller than the 
recoil energy and the dispersion relation should be given by the tight-binding approximation.
Another way to check that the complexes will not be severely modified by the presence of the 
lattice for small $s$, is to consider the bandwidth of the lowest lattice band, $W_0$. It 
is given by $W_0/E_0=\tfrac{16}{\sqrt{\pi}}s^{3/4}e^{-2\sqrt{s}}$ \cite{pethick2008}.
For $s=5$ we find $W_0/E_0\sim 0.345$ or $W_0=1.7\tfrac{\hbar^2}{m\Delta^2}$. The latter
is comparable to the binding energy for $U_0\sim 2.5$ and we therefore do not expect 
strong modification of the bound state energies and structures.

Finally, we address the external parabolic confinement that is always present in experiments with
ultracold atoms and molecules. This is most often approximated by an harmonic oscillator
in the plane or tube which is preferably shallow to approximate the homogeneous case. 
For a trapping angular frequency of $\omega$, then the quantity of interest is 
$\eta=E_{\text{trap}}/\tfrac{\hbar^2}{m\Delta^2}=\tfrac{m\Delta^2\omega}{\hbar}$. If $\eta$
is small, we expect the outer confinement to be negligible. In the recent 
experiments at JILA \cite{miranda2011} the in-plane external trap has a frequency of 36 Hz, 
yielding $\eta\sim 0.13$. The external trap is therefore expected to have small influence
on the bound states. In the opposite limit of large $\eta$, the bound complexes must be 
calculated with the external trap included. The methods used will work in that setup 
also.

We have suggested an optical detection scheme, which promises a less destructive, 
in situ measurement procedure up to a physically exciting QND limit. The few-body 
complexes can be also observed by other techniques including destructive ones. 
For example, the 1D lattice depth can be changed periodically. If the shaking 
frequency matches the binding energy of a complex it dissociates thereby heating the 
system. The temperature after shaking shows resonances as a function of frequency \cite{stroh2010}. 
Alternatively, RF spectroscopy can be used \cite{rf1,rf2}.

\section{Conclusions and outlook}
In this work we analyzed few-body complexes formed of dipolar molecules confined in two one-dimensional tubes.
 We studied the stability of these few-body complexes as a function of dipolar moment,
transverse confinement, tilting angle, and statistics of the molecules.
The complexes extend over both tubes and are bound by intertube attraction arising from the long-ranged dipolar interaction.
The extension of the complexes allows for an optical, non-destructible detection scheme, which enables in-situ probing of the complexes.
We show that the intensity of the scattered light jumps whenever a new complex becomes stable but stays constant otherwise.

Our work can be extended in various ways. As discussed in Sec.~\ref{sec2D} we expect similar results for bilayer systems instead of bitubes.
Another extension concerns possible few-body states for a larger number of tubes/layers. In the case of more than two adjacent tubes or layers, more complex few-body states become possible. In particular, chains of molecules can form with one molecule per tube/layer \cite{wang2006}.
By tilting the direction of the dipoles more than one molecule per layer can be bound which can lead to a bifurcation of the tubes.

In Sec.~\ref{many-body} we addressed the impact of our results on the many-body systems with a finite density in each tube.
Particularly interesting are fermionic systems. There the analog of the dimer physics in the many-body context is the BEC-BCS crossover at finite density \cite{zinner2010}. In the BEC regime the effective entities of the systems are bosonic and correspond to tightly bound dimers, while in the BCS regime the attraction leads to Cooper pairing but still the Fermi-energy is larger than the binding energy of the dimer state.
However, such a crossover will be modified by the presence of trimers.

We found for the few-body states an extended regime, where trimer are stable but dimer do not yet bind (forming the 2-2 complex). Translated to the many-body regime, this suggests that few-body complexes like trimers could be relevant for an intermediate interaction regime before the system undergoes Wigner crystallization in the strong interaction limit \cite{buchler2007}.
These examples of many-body physics are relevant for Fermi gases close to degeneracy, where the temperature is of the order of or smaller than the Fermi energy. However, we note that for strong dipole interaction stable complexes will be present even in experiments in thermal gases away from degeneracy. In particular the proposed detection scheme should also work on thermal clouds.

\section{Acknowledgements}
Special thanks to D. Pekker, J. von Stecher, M. Lukin, S. Rittenhouse, and M. Valiente.  Funding by DFG
grant WU 609/1-1, FWF project J3005-N16, and EPSRC Project EP/I004394/1, and by the Danish
Council for Independent Research $\vert$ Natural Sciences
is appreciated. The authors acknowledge support from the Army Research Office with funding from the DARPA OLEprogram, CUA, NSF Grant No. DMR-07-05472, AFOSR Quantum Simulation MURI, AFOSR MURI on Ultracold Molecules, and the ARO-MURI on Atomtronics.

\appendix

\section{Relative coordinates and space grid}\label{RelCoord}
The center of mass motion is unaffected by interactions and can be separated by introducing relative coordinates.

\begin{eqnarray}
&H=H_{\text{CM}}(u_0)+H_{\text{Rel}}(\{u\})&\\
&H_{\text{CM}}(u_0)=-\frac{\hbar^2}{2m\Delta^2}\partial_{u_0}^2&\notag\\
&H_{\text{Rel}}(\{u\})=\frac{\hbar^2}{m \Delta^2}\left(\sum_{j=1}^{N-1} \partial^2_{u_j}+U_0 V(\{u\})\right),&\notag
\end{eqnarray}
where $\{u\}=(u_1,...,u_{N-1})$ collectively denotes the $N-1$ relative coordinates.
There is freedom in choosing relative coordinates and we choose a transformation matrix, $M$, 
\begin{eqnarray*}
\left(\begin{array}{c} u_0\\..\\u_{N-1}\end{array}\right)&=&M\left(\begin{array}{c} x_0\\..\\x_{N-1}\end{array}\right)\\
\end{eqnarray*}
such that the kinetic energy is proportional to the unit matrix.
We chose the following forms of $M$ relating $\{u\}$ with the original coordinates for the different few-body complexes
\begin{eqnarray}
M_{D}&=&1/\sqrt{2} \left(\begin{array}{c c} 1 & 1 \\ -1 & 1 \end{array}\right)\label{eq:relD}\\[2ex]
M_{Tr}&=&\left(\begin{array}{c c c} 1/\sqrt{3} & 1/\sqrt{3} & 1/\sqrt{3}\\ -2/\sqrt{6} & 1/\sqrt{6} & 1/\sqrt{6}\\0 & -1/\sqrt{2} & 1\sqrt{2} \end{array}\right)\\[2ex]
M_{2-2}&=&\left(\begin{array}{c c c c}
 1/2 & 1/2 & 1/2 & 1/2\\
 1/2 & 1/2 & -1/2 & -1/2\\
 -1/\sqrt{2} & 1/\sqrt{2}& 0 & 0 \\
 0 & 0&-1/\sqrt{2}  & 1\sqrt{2}
 \end{array}\right)\label{eq:relTr}\\[2ex]
 M_{1-3}&=&\left(\begin{array}{c c c c}
 1/2 & 1/2 & 1/2 & 1/2\\
 -\sqrt{3} / {2} & 1/2 \sqrt{3} & 1/2 \sqrt{3} & 1/2 \sqrt{3}\\
 0 & \sqrt{2} / \sqrt{3}& -1/\sqrt{6} & 1/\sqrt{6} \\
 0 & 0&-1/\sqrt{2}  & 1/\sqrt{2}
 \end{array}\right)
\end{eqnarray}
The wave functions have the following symmetries:
\begin{eqnarray*}
&\psi_{Tr}(u_1,u_2)=s \psi_{Tr}(u_1,-u_2)&\\
&\psi_{2-2}(u_1,u_2,u_3)=s \psi_{F1}(u_1,-u_2,u_3)&\\
&=s \psi_{F1}(u_1,u_2,-u_3)&\\
&=p \psi_{2-2}(u_1,u_3,u_2)& \\
&\psi_{1-3}(u_1,u_2,u_3)=s \psi_{F2}(u_1,u_2,-u_3)&\\
&=s \psi_{F1}(u_1,\frac{-u_2+\sqrt{3} u_3}{2}, \frac{\sqrt{3}u_2+ u_3}{2})&\\
&=s \psi_{F1}(u_1,\frac{-u_2-\sqrt{3} u_3}{2}, \frac{-\sqrt{3}u_2+ u_3}{2})&
\end{eqnarray*}

Corresponding to the symmetries we use a square and cubic lattice for trimer and 2-2, while for 1-3 we choose a hexagonal lattice in the $u_2-u_3$ plane and a regular grid in $u_1$ plane. The 2-2 ground state will have the symmetry $p=1$.

The Laplace operator in square lattices with lattice constants $a_1$ and $a_2$ takes the form 
\begin{align}
\Delta f({\bf x})=\sum_{i=1}^2 \left[f({\bf x}+a_i \hat{e}_i)-2f({\bf x})+f({\bf x}-a_i \hat{e}_i)\right]/a_i^2
\end{align}
with $\hat{e}_1=(1,0)$
and $\hat{e}_2=(0,1)$. In the hexagonal lattice with lattice constant $a$ it is given by  
\begin{align}
\Delta f({\bf x})=\frac{2}{3} \sum_{i=1}^3 \left[f({\bf x}+a \hat{e}_i)-2f({\bf x})+f({\bf x}-a \hat{e}_i)\right]/a^2
\end{align}
with $\hat{e}_1=(-1,0)$, where $\hat{e}_2=\frac{1}{2}(\sqrt{3},-1)$ and $\hat{e}_2=\frac{1}{2}(\sqrt{3},1)$.

\end{document}